\documentclass{IEEEtaes}

\usepackage{color,array,amsthm}
\usepackage{graphicx}

\usepackage{amssymb}
\usepackage{booktabs}
\usepackage{upgreek}
\usepackage{cite}
\usepackage{amsmath,amsfonts}
\usepackage{bm}
\usepackage{siunitx}
\usepackage[caption=false,font=footnotesize,labelfont=rm,textfont=rm]{subfig}

\jvol{XX}
\jnum{XX}
\jmonth{XXXXX}
\paper{1234567}
\pubyear{2022}
\doiinfo{TAES.2022.Doi Number}

\begin{document}

\title{Waveform-Domain Complementary Signal Sets for Interrupted Sampling Repeater Jamming Suppression}

\author{Hanning Su}
\affil{National University of Defense Technology, Changsha 410073, China} 

\author{Qinglong Bao}
\affil{National University of Defense Technology, Changsha 410073, China} 

\author{Jiameng Pan}
\affil{National University of Defense Technology, Changsha 410073, China}

\author{Fucheng Guo}
\affil{National University of Defense Technology, Changsha 410073, China}

\author{Weidong Hu}
\affil{National University of Defense Technology, Changsha 410073, China}

\receiveddate{Manuscript received XXXXX 00, 0000; revised XXXXX 00, 0000; accepted XXXXX 00, 0000.}

\corresp{{\itshape (Corresponding author: Hanning Su and Qinglong Bao)}.}

\authoraddress{Hanning Su, Qinglong Bao, Jiameng Pan, Fucheng Guo and Weidong Hu are with the College of Electronic Science and Technology, National University of Defense Technology, Changsha 410073, China 
(e-mail: \href{mailto:hanningsu18@nudt.edu.cn}{hanningsu18@nudt.edu.cn}; \href{mailto:baoqinglong@nudt.edu.cn}{baoqinglong@nudt.edu.cn}; \href{mailto:panjiameng@nudt.edu.cn}{panjiameng@nudt.edu.cn}; \href{mailto:gfcly@21cn.com}{gfcly@21cn.com}; \href{mailto:wdhu@nudt.edu.cn}{wdhu@nudt.edu.cn}).}

\editor{This work was supported by the National Science Foundation of China under Grant 62231026.}

\markboth{AUTHOR ET AL.}{SHORT ARTICLE TITLE}
\maketitle

\begin{abstract}The interrupted-sampling repeater jamming (ISRJ) is coherent and has the characteristic of suppression and deception to degrade the radar detection capabilities. The study focuses on anti-ISRJ techniques in the waveform domain, primarily capitalizing on waveform design and and anti-jamming signal processing methods in the waveform domain. By exploring the relationship between waveform-domain adaptive matched filtering (WD-AMF) output and waveform-domain signals, we demonstrate that ISRJ can be effectively suppressed when the transmitted waveform exhibits waveform-domain complementarity. We introduce a phase-coded (PC) waveform set with waveform-domain complementarity and propose a method for generating such waveform sets of arbitrary code lengths. The performance of WD-AMF are further developed due to the designed waveforms, and simulations affirm the superior adaptive anti-jamming capabilities of the designed waveforms compared to traditional ones. Remarkably, this improved performance is achieved without the need for prior knowledge of ISRJ interference parameters at either the transmitter or receiver stages.
\end{abstract}


\section{INTRODUCTION}

I{\scshape nterrupted-sampling} repeater jamming (ISRJ) is a form of coherent jamming, capable of swiftly and accurately generating jamming signals using digital radio frequency memory (DRFM) \cite{1,2}. The jamming signals retransmitted by interference machines based on DRFM devices are coherent with the radar transmission signals, leading to the presence of both genuine and false target peaks in the range profile acquired through matched filtering. To suppress ISRJ and bolster radar detection capabilities, the anti-ISRJ methodology is studied in this work.

The leading theory concerning ISRJ was initially proposed by Wang et al. \cite{2}, attracting considerable attention and extensive research over more than a decade. In \cite{3} and \cite{4}, ISRJ was employed to create multiple false targets in the range profile, with an analysis of the characteristics of these false targets. To generate false targets ahead of the true target, ISRJ was enhanced by modulating the ISRJ signal using frequency-shifting jamming in \cite{5}. Non-periodic ISRJ was explored in \cite{6}, capable of producing a comprehensive jamming coverage over the radar's true target. In \cite{7}, a linear weighted optimization approach was adopted to rectify the uneven energy distribution of non-periodic ISRJ, resulting in the creation of a square jamming strip with more effective coverage.

As per publicly accessible literature, it is evident that existing techniques for mitigating ISRJ encompass both receiver-side signal processing methods and transmitter-side waveform design strategies. In the realm of receiver-side signal processing, ISRJ-induced false targets have been effectively eliminated through the skillful application of time-frequency analysis and band-pass filtering \cite{8,9,10,11}. The critical parameters associated with ISRJ have been deduced using time-frequency analysis and deconvolution processing. In the context of transmitter-side waveform design methodologies, certain researchers have sought to disrupt the Doppler continuity of interfering signals by constructing sparse Doppler waveforms \cite{12}, thereby facilitating the identification and suppression of interfering signals. Despite endeavors such as mismatch filtering \cite{13} and complementary waveforms \cite{14} to jointly optimize the design of transmission waveforms and mismatch filters through algorithmic means, the fundamental nature of these approaches still hinges on prior knowledge of segments related to the jamming signals. 

It is not arduous to discern that existing means of countering ISRJ exhibit a pronounced reliance on pattern recognition of jamming and the estimation of pivotal jamming parameters. Consequently, the primary step in most anti-jamming measures often involves estimating ISRJ key parameters through cognitive methods \cite{15,16,17,18,19}. These approaches, however, exhibit a certain degree of constraint in their adaptability to complex interference scenarios since they hinge upon the accuracy of cognitive models. This restriction impedes their further practical applicability in real-world interference scenarios. Furthermore, certain scholars have sought to employ deep neural networks for the extraction of jamming-free segments and the subsequent generation of band-pass filters \cite{20}. This adaptive approach to some extent diminishes the necessity for prior ISRJ information. Nonetheless, the training of neural networks demands copious data, and the performance of networks trained with simulated data necessitates further validation through radar-measured data.

Additionally, in our own work \cite{21}, we put forth a waveform-domain adaptive matched filtering (WD-AMF) technique using linear frequency modulated (LFM) waveforms as an illustration. This method achieved effective ISRJ suppression in the absence of prior ISRJ information. However, it is important to note that LFM waveforms may not be the most optimal choice for WD-AMF. This is because when the real target echo is overwhelmed by ISRJ in the waveform domain, it results in a loss of the main lobe level and an increase in sidelobe level in the WD-AMF output.

Relative to our previous publication discussed in \cite{21}, this article delves deeper into the tenets of ISRJ suppression by devising the most suitable waveform for WD-AMF. Consistent with the principles we adhered to in our previous work, the waveform is fashioned without any antecedent knowledge of the ISRJ parameters. The contributions of this article can be succinctly summarized as follows.\vspace*{-12pt}
\begin{enumerate}
\item Through an analysis of the interaction between WD-AMF and the transmitted waveform, it becomes evident that the primary reason for the main lobe level loss and sidelobe level increase in the WD-AMF output is largely due to the absence of waveform-domain complementarity in the radar waveform. This complementarity can be effectively achieved by using phase-coded (PC) waveforms.
\item To achieve waveform-domain complementarity for the phase-coded (PC) waveform, a definition of waveform-domain complementary signal sets has been introduced, and their properties have been demonstrated. A method for generating waveform-domain complementary signal sets of arbitrary code lengths has been proposed. It has been proven that this complementary signal set retains its complementarity in the waveform domain, unaffected by Doppler effects. Furthermore, real target echo signals and ISRJ interference in the waveform domain can be treated as non-overlapping monocomponent signals.
\item An enhanced version of WD-AMF, improved with respect to waveform-domain complementary signal sets, has been developed to bolster anti-ISRJ performance. Simulation results validate the effectiveness of the proposed techniques, affirming that the waveform-domain complementary signal sets represent the optimal choice for WD-AMF.
\end{enumerate}

The subsequent sections of this article are structured as follows. In Section \uppercase\expandafter{\romannumeral2}, the fundamental aspects of ISRJ are elucidated, and an analysis is conducted to explore the relationship between WD-AMF and the waveform-domain signals. Section \uppercase\expandafter{\romannumeral3} is dedicated to the definition, properties, and generation methods of waveform-domain complementary signal sets. Section \uppercase\expandafter{\romannumeral4} provides the pertinent expressions for waveform-domain complementary signal sets in WD-AMF, along with the underlying anti-jamming principles. The simulations which validate the proposed techniques and present research findings, are detailed in Section \uppercase\expandafter{\romannumeral5}. Lastly, this article culminates in Section \uppercase\expandafter{\romannumeral6} with a brief discussion of the study's conclusions.

\section{PROBLEM FORMULATION}

In this section, we will commence by analyzing the model of ISRJ, followed by an analysis of the relationship between WD-AMF output and waveform-domain signals. We will then derive the necessary constraints for the optimal waveform adaptation to WD-AMF.

\subsection{ISRJ model}

Assuming $s(t)$ represents the transmitted waveform and $\jmath(t)$ represents the interference waveform, the echo signal $x(t)$ can be expressed as follows \cite{2}:
\begin{equation}
x(t)=A_s s\left(t-\tau_{\mathrm{s}}\right)+A_\jmath \jmath\left(t-\tau_\jmath\right)
\label{eq1}
\end{equation}
where, $A_s$ and $A_\jmath$ represent the amplitudes of the target echo signal and the interference signal, while $\tau_s$ denotes the delay of the target echo signal, and $\tau_\jmath$ signifies the delay of the jamming signal. If we consider $\jmath(t)$ as an example of interrupted sampling direct repeater jamming (ISDRJ), where the interrupted sampling frequency is $f_J = \frac{1}{T_J}$ and the jamming slice width is $T_\jmath$. Hence, the matched filtering output of $x(t)$ can be expressed as:
\begin{equation}
\begin{aligned} 
x_{o}(t) & =A_s \chi\left(t-\tau_s, 0\right)+\sum_{q=-Q}^{Q} A_\jmath A_{q} \chi\left(t-\tau_\jmath,-q f_J\right) 
\\
& =s_{\mathrm{o}}(t-\tau_s)+\jmath_{d}(t-\tau_\jmath)
\end{aligned}
\label{eq2}
\end{equation}
where, $\chi\left(t, f_{d}\right) = \int_{-\infty}^{+\infty} s(\tau) s^{*}(t+\tau) e^{j 2 \pi f_{d} \tau} d \tau$ represents the ambiguity function of $s(t)$, $A_{q} = T_{\jmath} f_{J} \operatorname{sinc}\left(\pi q f_{J} T_\jmath\right)$ denotes the modulation function of the harmonics, $Q$ represents the number of transitions between jammer acquisition and transmission modes. $s_{\mathrm{o}}(t) = A_{\mathrm{s}} \chi\left(t, 0\right)$ signifies the matched filtering output of the target echo signal, and $\jmath_{d}(t) = \sum\limits_{q=-Q}^{Q} A_\jmath A_{q} \chi\left(t,-q f_{J}\right)$ represents the matched filtering output of ISDRJ.

Eq. (\ref{eq2}) illustrates that the output of ISDRJ can equivalently be viewed as the superposition of multiple weighted target matched filtering outputs. The number of false targets in the range profile is determined by the Doppler tolerance of $s(t)$. When the Doppler tolerance of $s(t)$ is high, multiple false targets will be generated in the range profile.

When the interference system operates in an interrupted sampling repetitive repeater jamming (ISRRJ) mode, the matched filtering result of ISRRJ can be equivalently expressed as multiple time-domain shifts of ISDRJ. In the range profile, it exhibits multiple clusters of false targets, with the interference characteristics within each cluster resembling those in the ISDRJ mode. The matched filtering output of ISRRJ can be expressed as:
\begin{equation}
\jmath_{r}(t) = \sum_{p=0}^{P-1}\jmath_d(t-pT_\jmath)
\label{eq3}
\end{equation}
where $P$ denotes the number of repetitive repeater.

However, when the jamming system employs a interrupted sampling cyclic repeater jamming (ISCRJ) mode, due to the fact that different slices have identical transmission delays only during the initial relay, during the second and subsequent relays, their delays are delayed by $(q-1)T_\jmath$. Consequently, the pulse compression result will contain multiple clusters of false targets. The distribution characteristics of the false target clusters can similarly be described using ISDRJ. The matched filtering output of ISCRJ can be expressed as:
\begin{equation}
\jmath_{c}(t) = \sum_{q=0}^{Q-1}\jmath_d(t-qT_\jmath-qT_J)
\label{eq4}
\end{equation}

\subsection{Waveform-domain adaptive matched filtering}

In our previous work \cite{21}, we introduced an extended domain within the signal matched filtering process, referred to as the waveform domain. This domain represents the integral interval of the impulse response $h(\mu)$ of the matching filter, specifically, $\mu\in \mathrm{U}=\left[-\frac{T}{2},\frac{T}{2}\right]$, where $T$ denotes the waveform pluse width. The product of the signal $x(t-\mu)$ and the impulse response $h(\mu)$ through the matching filter is referred to as the waveform response function at time $t$:
\begin{equation}
\begin{aligned}
v^{(t)}(\mu) 
&= x(t-\mu)h(\mu)
\\
&= A_ss(t-\mu-\tau_s)h(\mu)+A_\jmath\jmath(t-\mu-\tau_\jmath)h(\mu)
\\
&= v_s^{(t-\tau_s)}(\mu) + v_\jmath^{(t-\tau_\jmath)}(\mu)
\end{aligned}
\label{eq5}
\end{equation}
where $v_s^{(t)}(\mu) = A_ss(t-\mu)h(\mu)$ denotes the waveform response function of $s(t)$, and $v_\jmath^{(t)}(\mu) = A_\jmath\jmath(t-\mu)h(\mu)$ denotes the waveform response function of $\jmath(t)$.

In the waveform domain, we have formulated an adaptive threshold function $\hat E^{(t)}(\mu)$ with the purpose of integrating $v^{(t)}(\mu)$ over the set containing effective integration elements $\mathrm{U}^{(t)}_s$, while masking the ineffective integration element set $\mathrm{U}^{(t)}_{\jmath}$. Simultaneously, the process adaptively compensates for the masked elements in the unbiased estimation $\hat v^{(t)}(\mu)$. This compensation is derived from a randomly selected continuous subset $\Psi^{(t)} \subseteq \mathrm{U}^{(t)}_{s}$, with a length matching that of $\mathrm{U}_{\jmath}^{(t)}$. This process is referred to as waveform-domain adaptive matched filtering (WD-AMF), and its output $z_{o}(t)$ can be articulated as follows:
\begin{equation}
\begin{array}{l}
z_{o}(t) = \int_{\mathrm{U}^{(t)}_{s}} v^{(t)}(\mu) \mathrm{d} \mu+\int_{\Psi^{(t)}} \hat v^{(t)}(\mu) \mathrm{d} \mu
\end{array}
\label{eq6}
\end{equation}

In WD-AMF, the sets $\mathrm{U}_s^{(t)}$ and $\mathrm{U}_\jmath^{(t)}$ can be obtained through the following equation:
\begin{subequations}
\begin{equation}
\mathrm{U}_{\jmath}^{(t)}=\left\{\mu\Big|| \hat{v}^{(t)}(\mu\pm \gamma \mathrm{d} \mu) \mid>\hat E^{(t)}(\mu)\right\}
\label{eq7a}
\end{equation}
\begin{equation}
\mathrm{U}_{s}^{(t)}=\mathrm{C}_{\mathrm{u}} \mathrm{U}_{j}^{(t)}
\end{equation}
\label{eq7b}
\end{subequations}
where $\gamma \mathrm{d} \mu$ denotes the protective interval.

\subsection{ISRJ suppression analysis for WD-AMF}
Eq. (\ref{eq6}) illustrates that the two components of $z_o(t)$ are interrelated. When $v^{(t)}(\mu)$ represents a non-periodic signal, $z_o(t)$ experiences a significant energy accumulation, exemplified by $z_o(\tau_s)$. Conversely, when $v^{(t)}(\mu)$ embodies a periodic signal, $z_o(t)$ does not attain substantial amplification. Specifically, when $|\tau_\jmath-\tau_s|<T$, and if $t\neq\tau_s$, $\mathrm{U}^{(t)}_s$ is no longer a continuous interval. As a result, the integration of $v^{(t)}(\mu)$ over $\mathrm{U}^{(t)}_s$ and $\Psi^{(t)}$ may no longer yield a periodic function. This situation could lead to a certain degree of energy accumulation in $z_o(t)$, potentially resulting in an elevation of side-lobe levels.

Furthermore, owing to the inclusion relation $\Psi \subseteq \mathrm{U}^{(t)}_{s}$, when $\left \| \mathrm{U}^{(\tau_s)}_{s} \right \| < \frac{T}{2}$, where $\left\|\cdot\right\|$ represents the length of the set, $z_o(\tau_s)$ will encounter a loss of energy from the genuine echo signal. Assuming ISRJ to be a form of self-defensive repeater interference, wherein sampling and repeater are temporally asynchronous. The jamming slice width to the interrupted sampling period ratio, is expressed as $\varepsilon = \frac{T_\jmath}{T_J}\leqslant \frac{1}{2}$. Under different interference scenarios, ISRJ displays a marked discrepancy in its duty cycle. For instance, in the context of ISDRJ, the duty cycle is given by $\eta = \varepsilon$. Conversely, in the case of ISRRJ, the duty cycle is expressed as $\eta = P\cdot\varepsilon$. In the scenario of ISCRJ, the duty cycle takes the form $\eta = \frac{(Q+1)}{2}\cdot \varepsilon$. If $\tau_s=\tau_\jmath$, when $\eta > \frac{1}{2}$, it occasionally verifies that $\left \| \mathrm{U}^{(\tau_s)}_{s} \right \| < \frac{T}{2}$, and this circumstance is comparatively feasible in the context of ISRRJ.

The majority of constant modulus waveforms in WD-AMF processing encounter the aforementioned circumstances, presenting an urgent challenge for WD-AMF's adaptability to diverse interference scenarios. 

From the analysis above, it becomes evident that the performance deterioration of WD-AMF results from the waveform-domain overlapping when $v^{(t)}(\mu)$ is mismatched. Therefore, the ideal waveform adaptation for WD-AMF should meet the following criteria:
\begin{equation}
v^{(t)}(\mu)
=
\left
\{\begin{array}{ll}
v_s^{(t-\tau_s)}(\mu), & t=\tau_s 
\\
v_\jmath^{(t-\tau_\jmath)}(\mu), & t=\tau_\jmath
\\
0, &\text{else} 
\end{array}
\right.
\label{eq8}
\end{equation}

Hence, this study, with an emphasis on the perspective of transmit waveform design, does not endeavor to solely resolve the issue through a single pulse waveform but seeks a combination of waveforms that can effectively suppress interference signals in the waveform domain. This enhancement is achieved through inter-pulse processing to bolster the effectiveness of ISRJ interference mitigation across a broader range of scenarios.

\section{WAVEFORM DOMAIN COMPLEMENTARY SIGNAL SETS}

Since Eq. (\ref{eq8}) can be seen as a constraint on the cross-correlation properties before signal integration, we can start by designing waveforms that meet the constraint conditions based on phase-coded (PC) waveforms with good cross-correlation properties.

\subsection{Definitions and characteristics of waveform domain complementary sequence sets}

Assuming the transmitted $D$-sets sequence is represented as $\mathbf{A}=\{\bm{a}_0, \bm{a}_1, \ldots, \bm{a}_i, \ldots, \bm{a}_{D-1}\}^{\mathrm{T}}$, where $\bm{a}_i=\{a_i(0), a_i(1), \ldots, a_i(n), \ldots, a_i(N-1)\}$, the following equation holds to satisfy Eq. (\ref{eq8}):
\begin{subequations}
\begin{equation}
b^{(m)}(n) = \sum_{i=0}^{D-1}a_{i}(n+m)a_{i}^{\ast}(n)
\label{eq9a}
\end{equation}
\begin{equation}
\begin{array}{l}
b^{(m)}(n) =\left\{
\begin{array}{cl}
D, 
& m = 0 
\\ 
0, 
& m\ne0
\end{array}\right.\end{array}
\label{eq9b}
\end{equation}
\label{eq9}
\end{subequations}
where $\bm{b}^{(m)} = \{b^{(m)}(0), b^{(m)}(1), \ldots, b^{(m)}(N-1)\}$ represents the waveform response sequence at time $m$. As per Eq. (\ref{eq9b}), it is evident that $\mathbf{A}$ is a column orthogonal matrix. Since for $m \neq 0$, all elements of $\bm{b}^{(m)}$ are zeros, we refer to this characteristic as waveform domain complementarity. A sequence set that satisfies waveform domain complementarity is termed a waveform domain complementary sequence set. Consequently, the following lemmas and theorems hold:

{\it Theorem 1}: The length $N$ of sequences $\bm{a}_i$ that satisfy Eq. (\ref{eq9}) does not exceed the number of sequence sets, represented as $D$.

{\it Proof}: As $\mathbf{A}$ constitutes a column orthogonal matrix, it ensues that the dimension of the column space of $\mathbf{A}$ coincides with its rank, designated as $N$.

Since the dimension of the row space of a matrix is equivalent to the dimension of its column space, the row space dimension of matrix $\mathbf{A}$ is also $N$.

The number of rows $D$ in a matrix must be greater than or equal to the dimension of its row space. Therefore, we have the inequality:
\begin{equation}
N \leqslant D
\label{eq10}
\end{equation}






{\it Lemma 1}: Let matrix $\mathbf{A}$ satisfies Eq. (\ref{eq9}), then $\mathbf{A}$ constitutes a set of complementary sequence.

{\it Proof}: Let $\bm{\psi}_{\bm{a}_i, \bm{a}_i}$ denote the autocorrelation function of the sequence $\bm{a}_i$, and let $\bm{\psi}_{\bm{a}_i, \bm{a}_i}(m)$ denote the $m$-th element in the sequence. 
\begin{equation}
\sum_{i=0}^{D-1}\bm{\psi}_{\bm{a}_i, \bm{a}_i}(m) = \sum_{n=0}^{N-1}b^{(m)}(n) = 0, \quad m\ne0
\label{eq11}
\end{equation}

Since when $\mathbf{A}$ satisfies complementarity in the waveform domain, it necessarily fulfills time domain complementarity. Considering that in radar systems, transmitted waveforms often adhere to constant modulus constraints, in our subsequent discussions, we will focus on the case where $\mathbf{A}$ is a symbolic matrix. 

Therefore, when $\mathbf{A}$ forms a waveform domain complementary sequence set, it inevitably satisfies the correlation properties of complementary sequence sets \cite{22}:

{\it Lemma 2}: A waveform domain complementary sequence set comprises an even number of sequences.

{\it Lemma 3}: If the length $N$ of binary sequences that compose a waveform domain complementary sequence set is odd, then the number $D$ of sequences must be a multiple of $4$.

\subsection{Generation of waveform domain complementary sequence sets}

There are currently many methods to generate binary column orthogonal matrices \cite{23}. In this question, we introduce a method to generate matrices of any size $2^{r+1}\times N$. 

Since $N\leqslant D$, we can first generate a $D\times D$ Walsh-Hadamard matrix \cite{22}, and then select any $N$ columns to form a $D\times N$ waveform domain complementary sequence set $\mathbf{A}$.

A $D\times D$ Walsh-Hadamard matrix can be extended to have longer lengths through two typical iterative expansion methods: interleaving and cascading. This paper briefly introduces the cascading method \cite{22}. The following operations are defined: ${\mathbf{S}}$${\mathbf{S}}$ represents the cascading operation on ${\mathbf{S}}$, $\tilde{\mathbf{S}}$ represents the operation of reversing the row sequence of $\tilde{\mathbf{S}}$, and $-\mathbf{S}$ denotes the negation operation on $\mathbf{S}$.

To carry out the initial step of the recursive procedure, we construct first an orthogonal set $(\mathbf{s}_0, \mathbf{s}_1), D=2$, (see Golay \cite{24}). Next, the initial matrix $\Delta$ is constructed:
\begin{equation}
\Delta=\left[\begin{array}{rr}\mathbf{s}_{0} & \tilde{\mathbf{s}}_{1} \\ \mathbf{s}_{1} & -\tilde{\mathbf{s}}_{0}\end{array}\right]
\label{eq12}
\end{equation}

If we consider the matrix $\Delta$ as a three-dimensional matrix, where each column contains various binary sequences, it forms an orthogonal sequence set. Applying the cascading method to the third dimension of matrix $\Delta$ as follows allows us to obtain a matrix ${\Delta}'$ where each column's binary sequences also form an orthogonal sequence set.
\begin{equation}
\Delta^{\prime}=\left[\begin{array}{cc}\Delta \Delta & (-\Delta) \Delta \\ (-\Delta) \Delta & \Delta \Delta\end{array}\right]
\label{eq13}
\end{equation}

By iterating this method $r$ times, the resulting matrix ${\Delta}'$ contains a total of $R = 2^{r+1}$ Walsh-Hadamard matrices, each of size $2^{r+1}\times2^{r+1}$. Therefore, we can choose any one of the Walsh-Hadamard matrices and select any $N$ columns from that matrix to obtain the waveform domain complementary sequence set $\mathbf{A}$.

\subsection{Waveform domain complementarity of the WDCSS waveform}

The waveform domain complementary sequence set necessitates the baseband modulation of each individual bit within the binary sequence to acquire the baseband waveform-domain complementary signal set, $S(t) = \{s_i(t)\}_{i=0}^{D-1}$, in the time domain:
\begin{equation}
s_{i}(t)=\sum_{n=0}^{N-1} a_{i}(n) \Omega\left(t-n T_{\mathrm{c}}\right)
\label{eq14}
\end{equation}
where $\Omega(t)$ denotes the baseband modulation signal, and $T_c$ represents the symbol duration. For the sake of clarity, in all subsequent sections, we shall refer to $S(t)$ as the WDCSS waveform.

Therefore, the waveform response function of the WDCSS waveform $S(t)$ can be expressed as:
\begin{equation}
w_s^{(t)}(\mu) = \sum_{i=0}^{D-1}s_{i}(t+\mu)s_{i}^{\ast}(\mu)
\label{eq15}
\end{equation}
and its numerical solution can be represented as:
\begin{equation}
\begin{array}{l}
w_s^{(t)}(\mu) 
\\
=\left\{
\begin{array}{cl}
\sum_{n=0}\limits^{N-1}D\cdot\Omega\left(t-nT_c+\mu\right)\Omega^{\ast}(\mu-nT_c), 
& |t| \leqslant T_c
\\
\\ 
0, 
& |t|>T_c
\end{array}\right.
\end{array}
\label{eq16}
\end{equation}

If there exists a Doppler component $e^{j2\pi f_d t}$, defining the $\left(\mu,f_d\right)$ plane at time $t$ as $\mho^{(t)}(\mu,f_d)=w^{(t)}(\mu)e^{j2\pi f_d \left(t+\mu\right)}$, substituting into Eq. (\ref{eq16}) yields:
\begin{equation}
\begin{array}{l}
\mho^{(t)}(\mu,f_d)
\\
=\left\{
\begin{array}{cl}
\sum_{n=0}\limits^{N-1}D\cdot&\Omega\left(t-nT_c+\mu\right)\Omega^{\ast}(\mu-nT_c)e^{j2\pi f_d \left(t+\mu\right)}, 
\\
& |t| \leqslant T_c
\\ 
0, 
\\
& |t|>T_c
\end{array}\right.
\end{array}
\label{eq17}
\end{equation}

Eq. (\ref{eq17}) indicates that when $|t| > T_c$, the Doppler shift cannot alter the complementary properties of the waveform set $S(t)$ in the waveform domain. 

If $\jmath(t)$ represents an example of JSDRJ, its waveform response function can be represented as:
\begin{equation}
w_{\jmath_d}(t) = g(t+\mu)w_s^{(t)}(\mu)
\label{eq18}
\end{equation}
where $g(t)$ denotes the intermittent sampling signal of the interference. And its numerical solution can be represented as:
\begin{equation}
\begin{array}{l}
w_{\jmath_d}^{(t)}(\mu) 
\\
=\left\{
\begin{array}{cl}
g(t+\mu)w_s^{(t)}(\mu), 
& |t| \leqslant T_c
\\
\\ 
0, 
& |t|>T_c
\end{array}\right.\end{array}
\label{eq19}
\end{equation}

Eq. (\ref{eq19}) indicates that when $\jmath(t)$ is ISDRJ, $w_{\jmath_d}^{(t)}(\mu)$ satisfies waveform domain complementarity.

If $\jmath(t)$ represents an example of JSRRJ, its waveform response function can be represented as:
\begin{equation}
w_{\jmath_r}^{(t)}(\mu) = \sum_{p=0}^{P-1}w_{\jmath_d}^{(t-pT_\jmath)}(\mu)
\label{eq20}
\end{equation}
if $T_\jmath\geqslant T_c$, its numerical solution can be represented as:
\begin{equation}
\begin{array}{l}
w_{\jmath_r}^{(t)}(\mu) 
\\
=\left\{
\begin{array}{cl}
w_{\jmath_d}^{(t-pT_\jmath)}(\mu), 
& |t-pT_\jmath| \leqslant T_c
\\
\\ 
0, 
& \text{else}
\end{array}\right.\end{array}
\label{eq21}
\end{equation}

Eq. (\ref{eq21}) indicates that when $\jmath(t)$ is ISRRJ, and $T_\jmath\geqslant T_c$, $w_{\jmath_r}^{(t)}(\mu)$ can be considered as a linear shift of $w_{\jmath_d}^{(t)}(\mu)$, and at each moment $t = pT_\jmath$, the signal duty cycle in the waveform domain is given by $\eta=\varepsilon$.

If $\jmath(t)$ represents an example of JSCRJ, its waveform response function can be represented as:
\begin{equation}
w_{\jmath_c}^{(t)}(\mu) = \sum_{q=0}^{Q-1}w_{\jmath_d}^{(t-qT_\jmath-qT_J)}(\mu)
\label{eq22}
\end{equation}
if $T_\jmath+T_J\geqslant T_c$, its numerical solution can be represented as:
\begin{equation}
\begin{array}{l}
w_{\jmath_c}^{(t)}(\mu) 
\\
=\left\{
\begin{array}{cl}
w_{\jmath_d}^{(t-qT_\jmath-qT_J)}(\mu), 
& |t-qT_\jmath-qT_J| \leqslant T_c
\\
\\ 
0, 
& \text{else}
\end{array}\right.\end{array}
\label{eq23}
\end{equation}

Eq. (\ref{eq23}) indicates that when $\jmath(t)$ is ISCRJ, and $T_\jmath+T_J\geqslant T_c$, $w_{\jmath_r}^{(t)}(\mu)$ can be considered as a linear shift of $w_{\jmath_d}^{(t)}(\mu)$, and at each moment $t = qT_\jmath+qT_J$, the signal duty cycle in the waveform domain is given by $\eta=\varepsilon$.

Therefore, when the time delay difference between the ISRJ and the true echo signal, $|\tau_{\jmath}-\tau_{s}| > T_\jmath \geqslant T_c$, the following equation holds:
\begin{subequations}
\begin{equation}
w^{(t)}(\mu) = w_{s}^{(t-\tau_s)}(\mu) + w_{\jmath}^{(t-\tau_{\jmath})}(\mu)
\label{eq24a}
\end{equation}
\begin{equation}
\begin{array}{l}
w^{(t)}(\mu)=\left\{
\begin{array}{cl}
w_{s}^{(t-\tau_s)}(\mu), & |t-\tau_s|\leqslant T_c
\\ 
w_{\jmath}^{(t-\tau_{\jmath})}(\mu), & |t-\tau_{\jmath}|\leqslant T_c
\\
0, & \text{else.}
\end{array}\right.
\end{array}
\label{eq24b}
\end{equation}
\label{eq24}
\end{subequations}

Eq. (\ref{eq24}) demonstrates that when $|\tau_{\jmath}-\tau_{s}| > T_\jmath \geqslant T_c$, the non-zero elements of $w_{s}^{(t-\tau_s)}(\mu)$ and $w_{\jmath}^{(t-\tau_{\jmath})}(\mu)$ do not overlap at any time $t$. This ensures that $w_{s}^{(t-\tau_s)}(\mu)$ and $w_{\jmath}^{(t-\tau_{\jmath})}(\mu)$ do not cross interference. This characteristic of waveform domain complementarity provides significant convenience for subsequent waveform domain processing.

\section{THE WD-AMF FOR THE WDCSS WAVEFORM}

As per the analysis from the previous section, when the condition $|\tau_{\jmath}-\tau_{s}| > T_\jmath \geqslant T_c$ is satisfied, all interference signals generated by different ISRJ modes can be considered as standard ISDRJ signals with different delays in the waveform domain. Therefore, we only need to discuss how to suppress ISDRJ using the WDCSS waveform and WD-AMF, which can handle all modes of ISRJ. Thus, this section will investigate the suppression of ISDRJ through WD-AMF and the WDCSS waveform.

\subsection{Waveform domain adaptive threshold function}

To describe the variation of $w^{(t)}(\mu)$ during the waveform domain integration process, we define the cumulative waveform coherence function as:
\begin{equation}
y^{(t)}(\rho)=\int_{-\infty}^{\rho} w^{(t)}(\mu) \mathrm{d} \mu
\label{eq25}
\end{equation}

It is evident that when $\rho \rightarrow \infty$, $y^{(t)}(\rho)$ corresponds to the output of the matched filter at time $t$.

Combining Eq. (\ref{eq16}), (\ref{eq19}), (\ref{eq24}) and (\ref{eq25}), we obtain:
\begin{equation}
\begin{array}{l}
y^{(t)}(\rho)=\left\{
\begin{array}{cl}
y_{s}^{(t-\tau_s)}(\rho), & |t-\tau_s|\leqslant T_c
\\ 
y_{\jmath}^{(t-\tau_{\jmath})}(\rho), & |t-\tau_{\jmath}|\leqslant T_c
\\
0, & \text{else.}
\end{array}\right.
\end{array}
\label{eq26}
\end{equation}
where $y_s^{(t)}(\rho)=\int_{-\infty}^{\rho} w_s^{(t)}(\mu) \mathrm{d} \mu$, and $y_\jmath^{(t)}(\rho)=\int_{-\infty}^{\rho} w_\jmath^{(t)}(\mu) \mathrm{d} \mu$. Especially, when $t=\tau_s$, or $t=\tau_\jmath$, we have:
\begin{subequations}
\begin{equation}
y^{(\tau_s)}(\rho)=y_{s}^{(0)}(\rho)=A_sD\left(\rho+\frac{T}{2}\right)
\label{eq27a}
\end{equation}
\begin{equation}
\begin{aligned}
y^{(\tau_\jmath)}(\rho)
&=y_{\jmath}^{(0)}(\rho)
\\
&=\begin{array}{l}
\left\{
\begin{array}{cl}
A_\jmath D\left(\rho+\frac{T}{2}-qT_J\right)+qDT_\jmath,
\\
\quad qT_J\leqslant\rho-\frac{T}{2}\leqslant qT_J+T_\jmath
\\
\\
(q+1)A_\jmath DT_\jmath
\\
\quad qT_J+T_\jmath<\rho-\frac{T}{2}<(q+1)T_J
\end{array}\right.
\end{array}
\end{aligned}
\label{eq27b}
\end{equation}
\label{eq27}
\end{subequations}

Eq. (\ref{eq27}) indicates that $y^{(\tau_s)}(\rho)$ is a linear function, implying that the energy accumulation of $w^{(\tau_s)}(\mu)$ in the waveform domain grows linearly. On the other hand, $y^{(\tau_\jmath)}(\rho)$ is a piecewise linear function, indicating that the energy accumulation of $w^{(\tau_\jmath)}(\mu)$ in the waveform domain grows in a piecewise linear manner. Hence, we can define a linear objective function $O^{(t)}(\rho)$ to characterize the average linear growth process of $y^{(t)}(\rho)$. 
\begin{equation}
O^{(t)}(\rho) = o^{(t)}\rho
\label{eq28}
\end{equation}
where $o^{(t)}$ represents the average rate of energy growth for $y^{(t)}(\rho)$. As the matched filter is designed to maximize the output signal-to-noise ratio(SNR), the following equation must hold:
\begin{subequations}
\begin{equation}
o^{(t)} = \frac{y^{(t)}\left(\frac{T}{2}\right)}{T}
\label{eq29a}
\end{equation}
\begin{equation}
\begin{array}{l}
\left\{
\begin{array}{cl}
o^{(t)}\leqslant A_sD=|w^{(t)}(\mu)|, \quad&|t-\tau_s|\leqslant T_c
\\
o^{(t)}\leqslant \varepsilon A_\jmath D=\varepsilon|w^{(t)}(\mu)|, \quad&|t-\tau_{\jmath}|\leqslant T_c
\\
0, \quad&\text{else}
\end{array}\right.
\end{array}
\label{eq29b}
\end{equation}
\label{eq29}
\end{subequations}

In Eq. (\ref{eq29b}), the equality in the inequality concerning $o^{(t)}$ holds if and only if the equality in the inequality concerning $t$ is satisfied. Eq. (\ref{eq29}) demonstrates a natural adaptive constraint relationship between the variable $o^{(t)}$ and $|w^{(t)}(\mu)|$. Therefore, an adaptive threshold function $E^{(t)}(\mu)$ can be designed to serve as the condition for distinguishing effective integration elements in the waveform domain $\mathrm{U}^{(t)}_s$ from ineffective integration elements $\mathrm{U}^{(t)}_\jmath$: 
\begin{subequations}
\begin{equation}
\mathrm{U}^{(t)}_s=\Big\{\mu\Big||w^{(t)}(\mu)|\leqslant E^{(t)}(\mu)\Big\}
\label{eq30a}
\end{equation}
\begin{equation}
\mathrm{U}^{(t)}_\jmath=\Big\{\mu\Big||w^{(t)}(\mu)|> E^{(t)}(\mu)\Big\}
\label{eq30b}
\end{equation}
\label{eq30}
\end{subequations}

Next, let's analyze the boundary conditions of the adaptive threshold function $E^{(t)}(\mu)$ that satisfy the decision criteria. If we set 
\begin{equation}
E^{(t)}(\mu) = \lambda \left|o^{(t)}\right|
\label{eq31}
\end{equation}
to ensure that $w^{(t)}(\mu)$ satisfies Eq. (\ref{eq30a}) when $|t-\tau_s|\leqslant T_c$, we must have $\lambda \geqslant 1$. Similarly, to ensure that $w^{(t)}(\mu)$ satisfies Eq. (\ref{eq30b}) when $|t-\tau_\jmath|\leqslant T_c$, we must have $\lambda \varepsilon \leqslant 1$. Therefore, the boundary conditions for $\lambda$ should be $1 \leqslant \lambda \leqslant \frac{1}{\varepsilon}$. Typically, $\varepsilon$ is unknown, so we can narrow down the boundary conditions to 
\begin{equation}
1 \leqslant \lambda = \frac{1}{\varepsilon_0} \leqslant \frac{1}{\varepsilon}
\label{eq32} 
\end{equation}
where $\varepsilon_0 \geqslant \varepsilon$. It is important to emphasize that $\varepsilon_0$ is an upper bound, meaning that $E^{(t)}(\mu)$ has adaptive decision capability for all interference elements with duty cycles less than $\varepsilon_0$. For example, in the case of self-jamming interference, where $\varepsilon \leqslant \frac{1}{2}$, $\varepsilon_0 = \frac{1}{2}$, and $\lambda = 2$.

\subsection{State estimation of waveform domain signals}

Eq. (\ref{eq29}), (\ref{eq30}) and (\ref{eq31}) indicate that in the presence of time domain noise, a necessary condition for ensuring algorithm robustness is to obtain unbiased estimates of $w^{(t)}(\mu)$ and $y^{(t)}(\frac{T}{2})$. 

Assuming that time domain noise has a mean of 0 and a variance of $\sigma^{2}$, due to the additivity property of Gaussian distributions, it is known that the Gaussian white noise $wgn^{(t)}(\mu)$ distributed on $w^{(t)}(\mu)$ within one coherently processed interval (CPI) follows the distribution characteristics:
\begin{equation}
wgn^{(t)}(\mu)\sim\left(0,D\sigma^2\right)
\label{eq33}
\end{equation}

As the indefinite upper limit integral of $wgn^{(t)}(\mu)$ is a Markov random process, the noise distributed on $y^{(t)}(\mu)$ is a Brownian noise $bn^{(t)}(\mu)$ that follows the distribution:
\begin{equation}
bn^{(t)}(\mu)\sim\left(0,\left(\mu+\frac{T}{2}\right)D\sigma^2\right)
\label{eq34}
\end{equation}

Considering the linear and piecewise linear relationships in Eq. (\ref{eq27}), we can obtain unbiased estimates, $\hat{y}^{(t)}(\frac{T}{2})$, and $\hat{w}^{(t)}(\mu)$ using the Interactive Multiple Model Kalman Filter (IMM-KF) algorithm \cite{25}. This relationship can be described by a linear model, $\hat{M}_1^{(t)}$, and two impulse models, $\hat{M}_2^{(t)}$ and $\hat{M}_3^{(t)}$:
\begin{equation}
\begin{aligned}
&\hat{\mathbf{M}}^{(t)}(\mu \mid \mu)
\\
&=u_{1} \hat{\mathbf{M}}_{1}^{(t)}(\mu \mid \mu)+u_{2} \hat{\mathbf{M}}_{2}^{(t)}(\mu \mid \mu)+u_{3} \hat{\mathbf{M}}_{3}^{(t)}(\mu \mid \mu)
\end{aligned}
\label{eq35}
\end{equation}
in which the weights $u_1$, $u_2$, and $u_3$ are ascertained for each model based on the residuals and residual covariance acquired through the implementation of the Kalman filter. $\hat{\mathbf{M}}^{(t)}_1(\mu|\mu)$, $\hat{\mathbf{M}}^{(t)}_2(\mu|\mu)$, and $\hat{\mathbf{M}}^{(t)}_3(\mu|\mu)$ denote the estimated state matrices of their respective models, and their one-step prediction state equations are delineated as follows:
\begin{subequations}
\begin{equation}
\begin{aligned}
&\hat{\mathbf{M}}_1(\mu+d\mu|\mu)
= \mathbf{F}_1\hat{\mathbf{M}}(\mu|\mu)
\\
&=\left[\begin{array}{ccccc}
1 & d\mu & 0 & 0 & 0\\  0 & 1 & 0 & 0 & 0\\ 0 & 0 & 0 & 0 & 0\\ 0 & 0 & 0 & 1 &0 \\ 0 & 0 & 0& 0 & 1
\end{array}\right]
\left[\begin{array}{l}
\hat y^{(t)}(\mu) \\ \hat w^{(t)}(\mu) \\ \hat \delta_{-}^{(t)}(\mu) \\ \hat \delta_{+}^{(t)}(\mu) \\ wgn^{(t)}(\mu) \end{array}\right]
\end{aligned}
\label{eq36a}
\end{equation}
\begin{equation}
\begin{aligned}
&\hat{\mathbf{M}}_2(\mu+d\mu|\mu)
=\mathbf{F}_2 \hat{\mathbf{M}}(\mu|\mu)
\\
&=\left[\begin{array}{ccccc}
1 & d\mu & d\mu & 0 & 0\\  0 & 1 & d\mu & 0 & 0\\ 0 & -1 & 0 & 0 & 0\\ 0 & 0 & 0 & 1 & 0\\ 0& 0 & 0 &0 & 1
\end{array}\right]
\left[\begin{array}{l}
\hat y^{(t)}(\mu) \\ \hat w^{(t)}(\mu) \\ \hat \delta_{-}^{(t)}(\mu) \\ \hat \delta_{+}^{(t)}(\mu) \\ wgn^{(t)}(\mu)\end{array}\right]
\end{aligned}
\label{eq36b}
\end{equation}
\begin{equation}
\begin{aligned}
&\hat{\mathbf{M}}_3(\mu+d\mu|\mu)
= \mathbf{F}_3 \hat{\mathbf{M}}^{(t)}(\mu|\mu)
\\
&=\left[\begin{array}{ccccc}
1 & d\mu & 0 & d\mu & 0\\  0 & 1 & 0 & d\mu & 0\\ 0 & 0 & 0 & 0 & 0\\ 0 & 0 & 0 & 1 & 0\\ 0 & 0 & 0 &0 &1
\end{array}\right]
\left[\begin{array}{l}
\hat y^{(t)}(\mu) \\ \hat w^{(t)}(\mu) \\ \hat \delta_{-}^{(t)}(\mu) \\ \hat \delta_{+}^{(t)}(\mu)\\wgn^{(t)}(\mu) \end{array}\right]
\end{aligned}
\label{eq36c}
\end{equation}
\label{eq36}
\end{subequations}
where $\mathbf{F}_i$, with $i=1,2,3$, signifies the matrices governing state transitions. The entities $\hat \delta_{-}^{(t)}(\mu)$ and $\hat \delta_{+}^{(t)}(\mu)$ pertain to distinct impulse functions, influencing both the direction and magnitude of $\hat w^{(t)}(\mu)$. Specifically, $\hat \delta_{-}^{(t)}(\mu+d\mu|\mu) = -\hat w^{(t)}(\mu)$ and $\hat \delta_{+}^{(t)}(\mu+d\mu|\mu) = \hat \delta_{+}^{(t)}(\mu) = Ko^{(t)}$, $K\geqslant \frac{1}{\varepsilon}$. It is noteworthy that the state matrix has been expanded in this context due to the measurement value $y^{(t)}(\mu)$ conforming to the Brownian noise model. And the probability transition matrix can be represented as:
\begin{equation}
\begin{aligned}
\mathbf{P}^{(t)} =
\left[\begin{array}{ccc}
1-2p_0 & p_0 & p_0\\  1 & 0 & 0 \\ 1 &0 &0
\end{array}\right]
\end{aligned}
\label{eq37}
\end{equation}
wherein, $p_0$ designates the likelihood of an abrupt alteration in the magnitude of $|w^{(t)}(\mu)|$, and the zero entries on the diagonal of the matrix are not rigorously zero but typically denote exceedingly minute values to guarantee the invertibility of the matrix.

Following the IMM-KF procedure, we can obtain unbiased estimates $\hat{w}^{(t)}(\mu)$ and $\hat{y}^{(t)}(\mu)$ for each time instant $t$. 

\subsection{The output of WD-AMF with the WDCSS waveform}

Upon obtaining the unbiased estimates $\hat{w}^{(t)}(\mu)$ and $\hat{y}^{(t)}(\mu)$, we can derive an unbiased estimate for the adaptive threshold function:
\begin{equation}
\hat{E}^{(t)} = \lambda \hat{o}^{(t)} = \lambda \cdot \frac{\hat{y}^{(t)}\left(\frac{T}{2}\right)}{T}
\label{eq38}
\end{equation}

Given that $w_\jmath^{(t)}(\mu)$ is not a continuous function, we must extend $\mathrm{U}_{\jmath}^{(t)}$. Subsequently, Eq. (\ref{eq30}) can be updated as follows:
\begin{subequations}
\begin{equation}
\mathrm{U}_{s}^{(t)}=\mathrm{C}_{\mathrm{u}} \mathrm{U}_{\jmath}^{(t)}
\label{eq39a}
\end{equation}
\begin{equation}
\mathrm{U}^{(t)}_\jmath=\Big\{\mu\Big||\hat{w}^{(t)}(\mu\pm \gamma \mathrm{d}\mu)|> \hat{E}^{(t)}\Big\}
\label{eq39b}
\end{equation}
\label{eq39}
\end{subequations}
where $\gamma\mathrm{d}\mu$ denotes the protective interval.

The output of WD-AMF at each time instant $t$ can then be expressed as the definite integral of $w^{(t)}(\mu)$ over $\mathrm{U}^{(t)}_s$:
\begin{equation}
w_o(t) = \int_{\mathrm{U}^{(t)}_s}w^{(t)}(\mu)d\mu
\label{eq40}
\end{equation}

Let's analyze the numerical solutions for $w_o(t).$ Since $\hat{E}^{(t)} \leqslant \lambda A_sD$ with $\lambda \geqslant 1$, let $t_\lambda$ denote the moment when $\hat{E}^{(t_\lambda)} = A_sD.$ When $|t - \tau_s| \leqslant |t_\lambda - \tau_s|$, taking into account the waveform domain complementarity described in Eq. (\ref{eq24b}), it is certain that $|w^{(t)}(\mu)| \leqslant \hat{E}^{(t_\lambda)}$, and in this case, $\mathrm{U}_s^{(t)}$ spans the entire waveform domain $\left[-\frac{T}{2},\frac{T}{2}\right]$. According to Lemma 1, we know that $S(t)$ is also a complementary signal set. Therefore, $t_\lambda$ is unique, and consequently, for all other time instants $t$, we have $\mathrm{U}^{(t)}_s=\emptyset$. Hence, the numerical solutions for $w_o(t)$ can be regarded as the numerical solutions for the main lobe width of the matched filter output $s_o(t)$ under a level of $-\frac{s_o(0)}{\lambda}$ below the main lobe peak, as follows:
\begin{equation}
\begin{array}{l}
w_o(t)=\left\{
\begin{array}{cl}
s_o(t-\tau_s), & |t-\tau_s|\leqslant |t_\lambda-\tau_s|
\\
0, & \text{else.}
\end{array}\right.
\end{array}
\label{eq41}
\end{equation}

To maintain robust sparsity in the distance domain of $w_o(t)$ and to facilitate subsequent target detection algorithms on $w_o(t)$, we typically apply compensation to the elements on $\mathrm{U}^{(t)}_\jmath$ equivalent to their original noise environment. Consequently, Eq. (\ref{eq40}) and Eq. (\ref{eq41}) can be further generalized as follows:
\begin{subequations}
\begin{equation}
w_o(t) = \int_{\mathrm{U}^{(t)}_s}w^{(t)}(\mu)d\mu + \int_{\mathrm{U}^{(t)}_\jmath}wgn^{(t)}(\mu)d\mu 
\label{eq42a}
\end{equation}
\begin{equation}
\begin{array}{l}
w_o(t)=\left\{
\begin{array}{cl}
s_o(t-\tau_s), & |t-\tau_s|\leqslant |t_\lambda-\tau_s|
\\
bn^{(t)}\left(\frac{T}{2}\right), & \text{else.}
\end{array}\right.
\end{array}
\label{eq42b}
\end{equation}
\label{eq42}
\end{subequations}

Through the previously described WD-AMF on the WDCSS waveform, complete suppression of ISRJ in any operational mode can be achieved.

\section{NUMERICAL EAMPLES}\label{5}

In this section, multiple simulations are provided to assess the proposed methodology. Initially, a numerical simulation is presented to substantiate the waveform domain complementarity and matched filtering performance of the WDCSS waveform. Subsequently, the proposed methods' effectiveness against anti-ISRJ is assessed in the presence of ISRJ under different operational modes. Finally, an analysis of the parametric sensitivity of the proposed approach is conducted. 

\subsection{Performance simulation of the WDCSS waveform}\label{5.A}
\begin{figure*}[htbp]
\centering
\subfloat[]{
\includegraphics[width=5.5 cm]{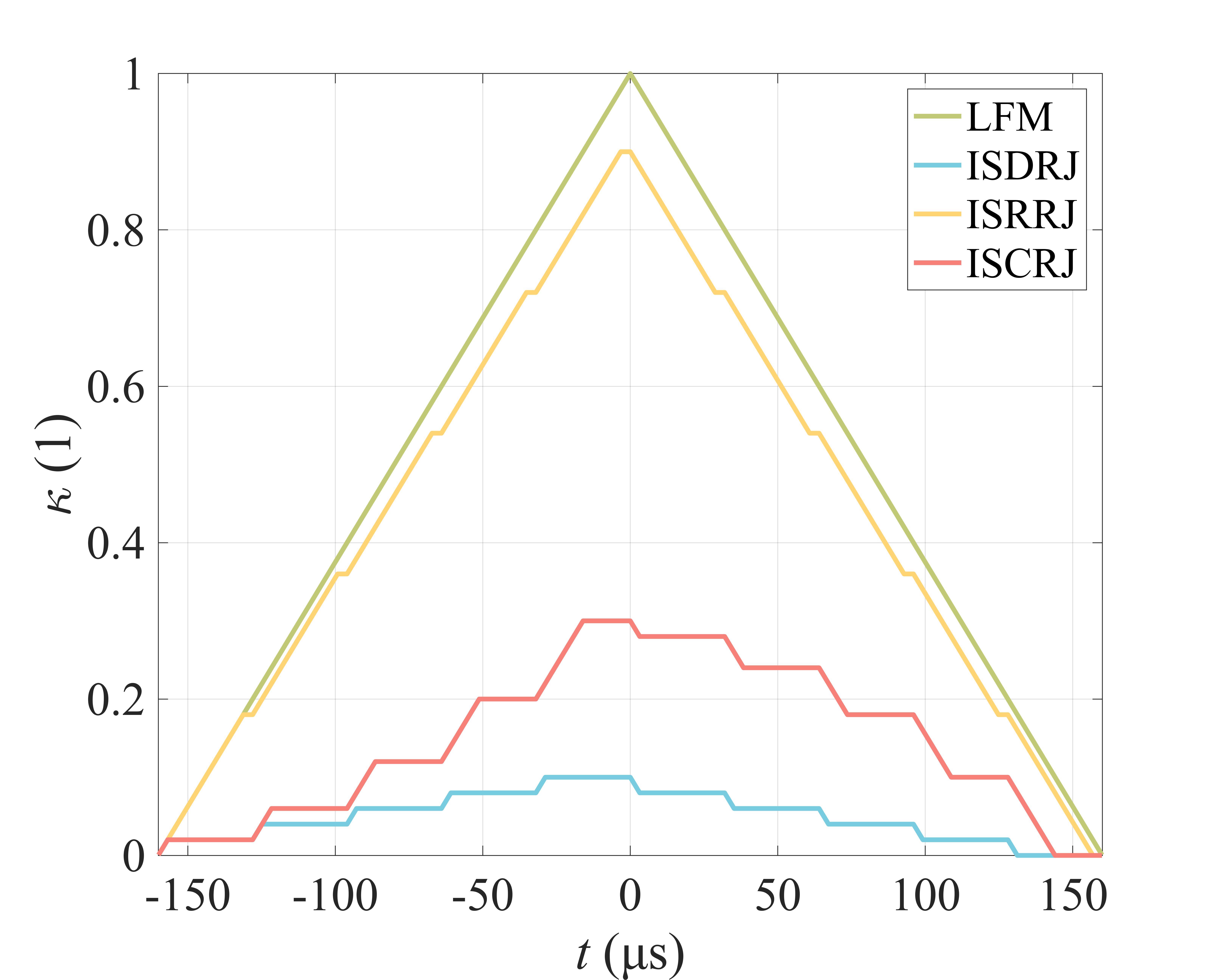}
\label{fig1(a)}}%
\subfloat[]{
\includegraphics[width=5.5 cm]{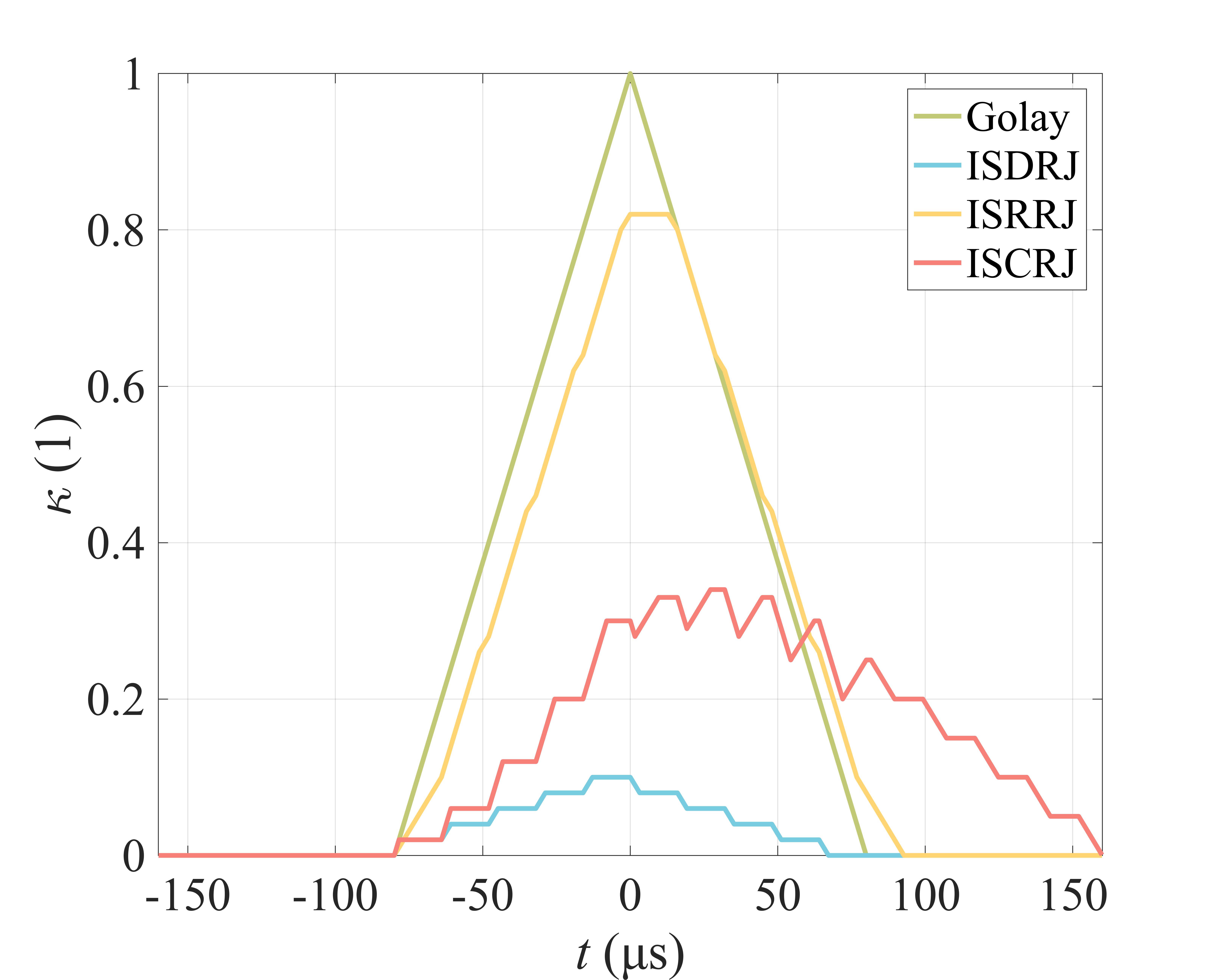}
\label{fig1(b)}}%
\subfloat[]{
\includegraphics[width=5.5 cm]{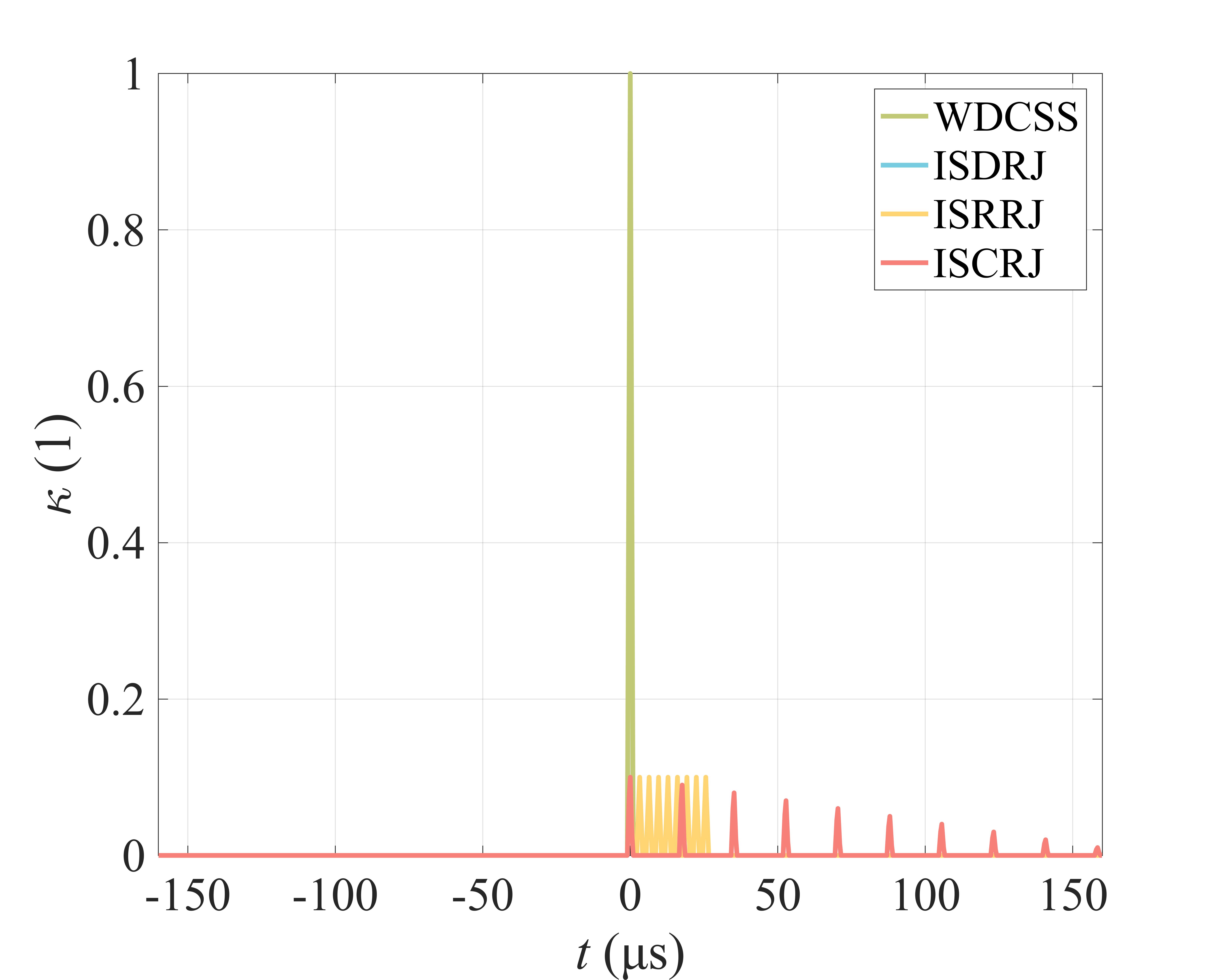}
\label{fig1(c)}}%
\centering
\caption{The variation of $\kappa$ with respect to time $t$ for different waveforms and their ISRJ characteristics. (a) LFM waveform; (b) Golay waveform; (c) WDCSS waveform.
\label{fig1}}
\end{figure*}

For radar, assuming a transmitted signal with a sampling frequency of $F_s=10$ MHz, and code length and number of pulses, $N=160$, $D=256$, the baseband modulation signal $\Omega(t)$ is a rectangular pulse, and the chip duration is $T_c = \SI{1}{\micro\second}$, hence the pulse width is $T = NT_c = \SI{160}{\micro\second}$. For the ISRJ jamming device, considering an intermittent sampling period of $T_{J} = \SI{32}{\micro\second}$ and a jamming
slice width to the interrupted sampling period ratio of $\varepsilon = 10\%$, the width of an individual jamming slice is thus $T_\jmath = \SI{3.2}{\micro\second}$. To delineate the complementarity of distinct signals in the waveform domain, we introduce $\kappa(t)$ as the ratio of the length of non-zero elements in $|w_s^{(t)}(\mu)|$ to the total duration $T$ in the waveform domain at time $t$:
\begin{equation}
\kappa(t) = \frac{\left\|\Big\{\mu\Big||w_s^{(t)}(\mu)|>0\Big\}\right\|}{T}
\end{equation}

Fig. \ref{fig1} illustrates the temporal evolution of $\kappa(t)$ concerning various waveforms and their corresponding operational modes under ISRJ conditions, where $P=9$ and $Q=5$. The comparative experiments involve three distinct signals: an LFM waveform featuring a $2$ MHz instantaneous bandwidth, a complementary Golay waveform with parameters $D=2$ and $N=160$, and the aforementioned WDCSS waveform. All three signals possess an identical pulse width.
\begin{figure}[htbp]
\includegraphics[width=6 cm]{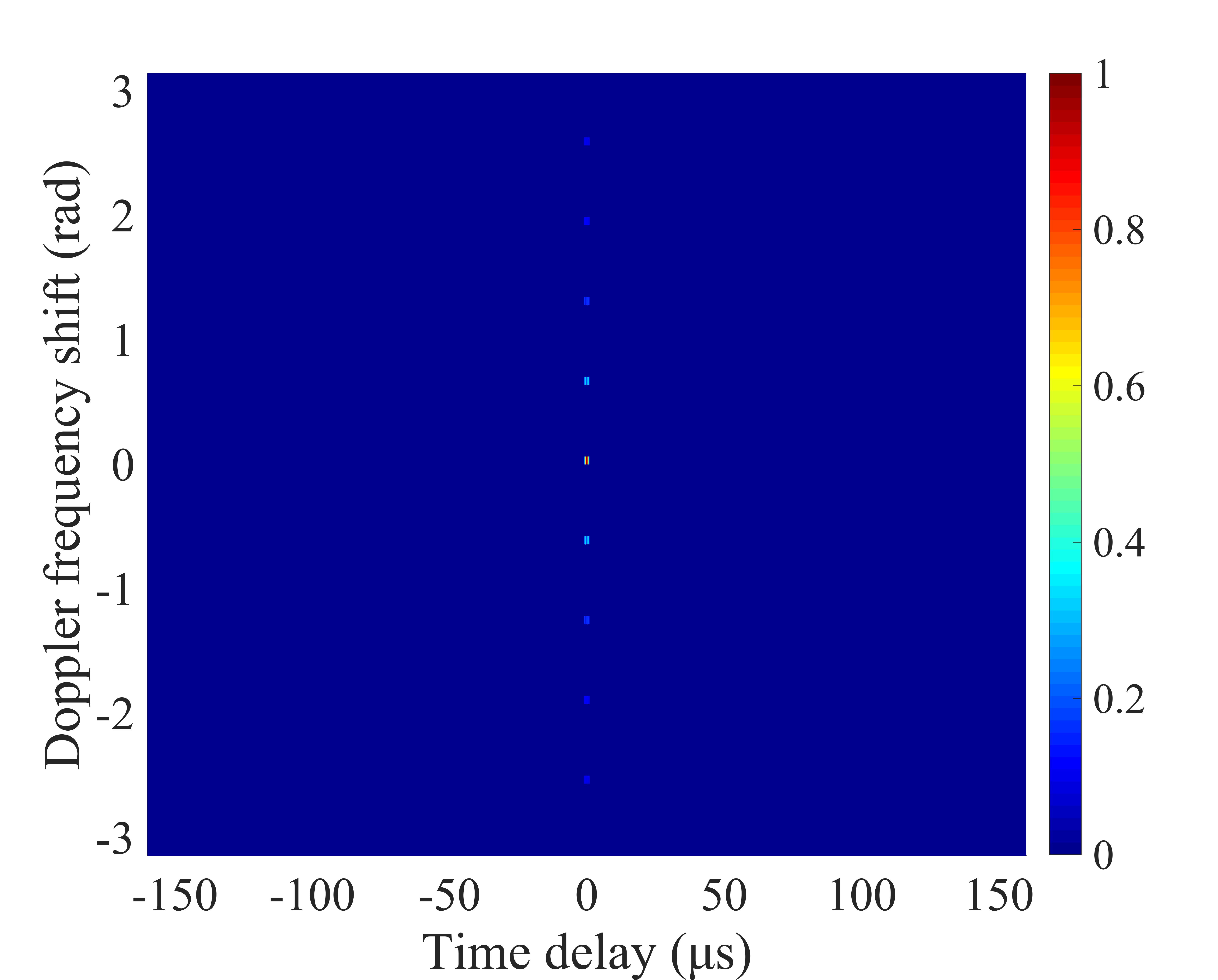}
\centering
\caption{The ambiguity function of the WDCSS waveform.
\label{fig2}}
\end{figure}

Simulation results reveal that for the LFM waveform and the Golay waveform, $\kappa(t)$ exhibits non-sparsity, leading to unavoidable overlapping between the target echo signal and the jamming signal in the waveform domain. It is evident that for these two waveforms, the jamming signal can occupy a duty cycle of up to $\eta$ in the waveform domain of the target echo signal. When $\eta>\frac{1}{2}$, the output energy of the WD-AMF signal for the target echo signal will be compromised.

However, for the WDCSS waveform, $\kappa(t)$ is sparse. The interference signal will occupy at most an $\varepsilon$ duty cycle of the waveform domain of the target echo signal. Specifically, as long as it satisfies $\tau_\jmath - \tau_s > T_c$, the target echo signal and the jamming signal in the waveform domain are completely orthogonal, and there will be no overlapping, regardless of the ISRJ mode and any $\eta$. In this case, the output energy of the target echo signal WD-AMF signal is always equal to its matched filter output, with no loss.

The ambiguity function of the WDCSS waveform, as shown in Fig. \ref{fig2}, reveals that the matched filter output of the WDCSS waveform is quite sensitive to Doppler frequency shifts. However, its complementary properties in the range profile are not sensitive to Doppler shifts.

\subsection{Performance evaluation for ISRJ resistance}
In this section, we will conduct simulations to validate the effectiveness of the WD-AMF method for the WDCSS waveform we have designed. Considering an L-band radar system and an interference system, the simulation parameters are as depicted in Tab. \ref{tab1}. The waveforms designed in Section \ref{5}-\ref{5.A} are employed herein, as their pulsewidth and bandwidth align with those discussed in Section \ref{5}-\ref{5.A}. The simulation scenario features a point target and an ISRJ jammer, and the positions of the targets and jammer are detailed in Tab. \ref{tab1}. The subsection simulates the ISRJ without modulation, and the input jamming-to-noise ratio (JNR) is set at $20$ dB for this simulation. ISRJ will be simulated in three different modes: ISDRJ, ISRRJ, and ISCRJ. All of them have the same $\varepsilon$ and $T_J$, and their parameters are also displayed in Tab. \ref{tab1}. It's important to note that in the simulation scenario as depicted in Tab. \ref{tab1}, it holds that $\tau_\jmath-\tau_s > T_\jmath > T_c$. Consequently, the WDCSS waveform exhibits waveform-domain complementarity. 

For comparative analysis, we present the WD-AMF \cite{21} outcomes, denoted as $z_o(t)$ (as referenced in Eq. (\ref{eq6})), when the radar employs LFM or Golay waveform sets. All waveforms possess an identical coherently processing interval, denoted as $\mathrm{CPI}=256$.
\begin{figure*}[htbp]
\centering
\subfloat[]{
\includegraphics[width=5.5 cm]{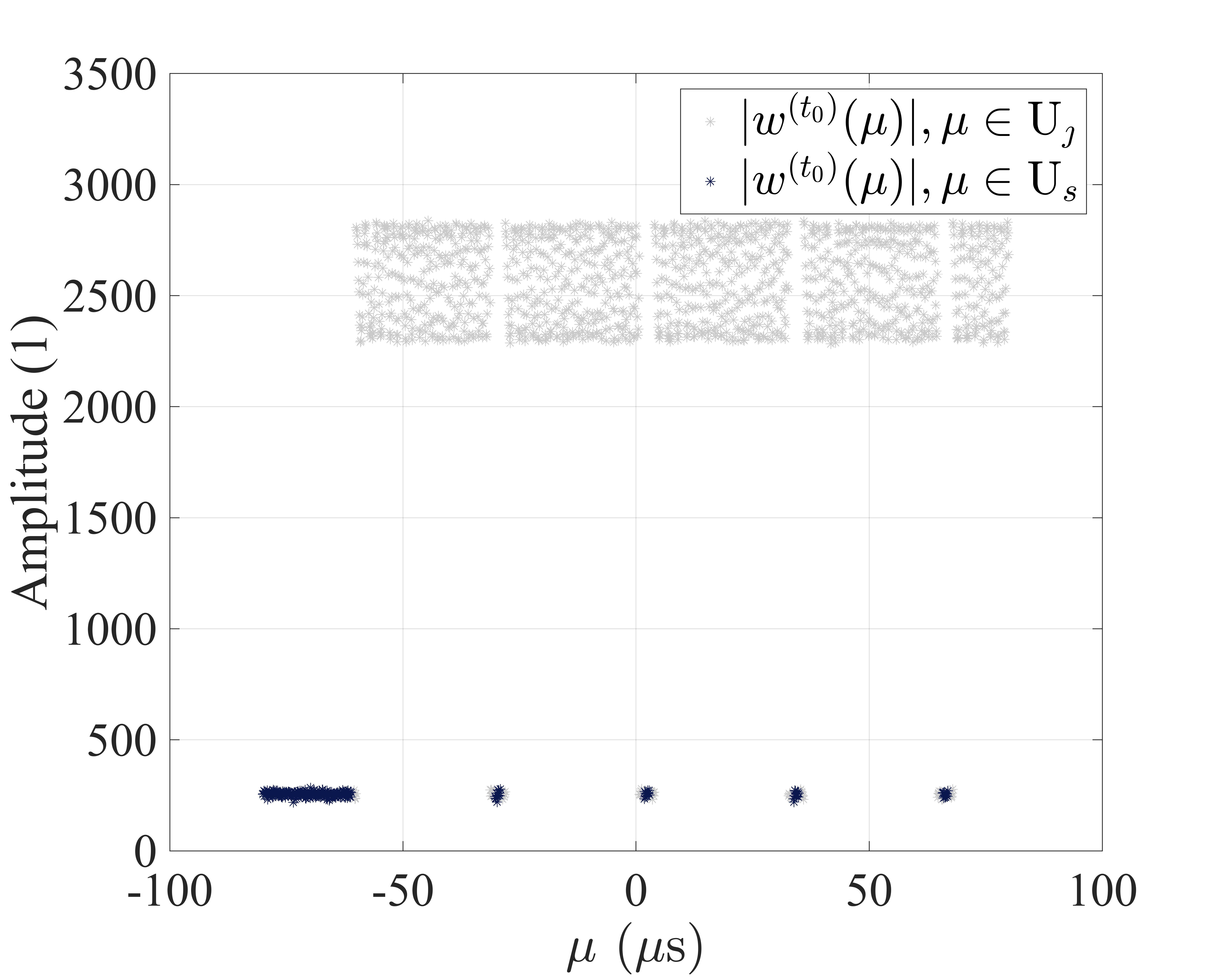}
\label{fig3(a)}}%
\subfloat[]{
\includegraphics[width=5.5 cm]{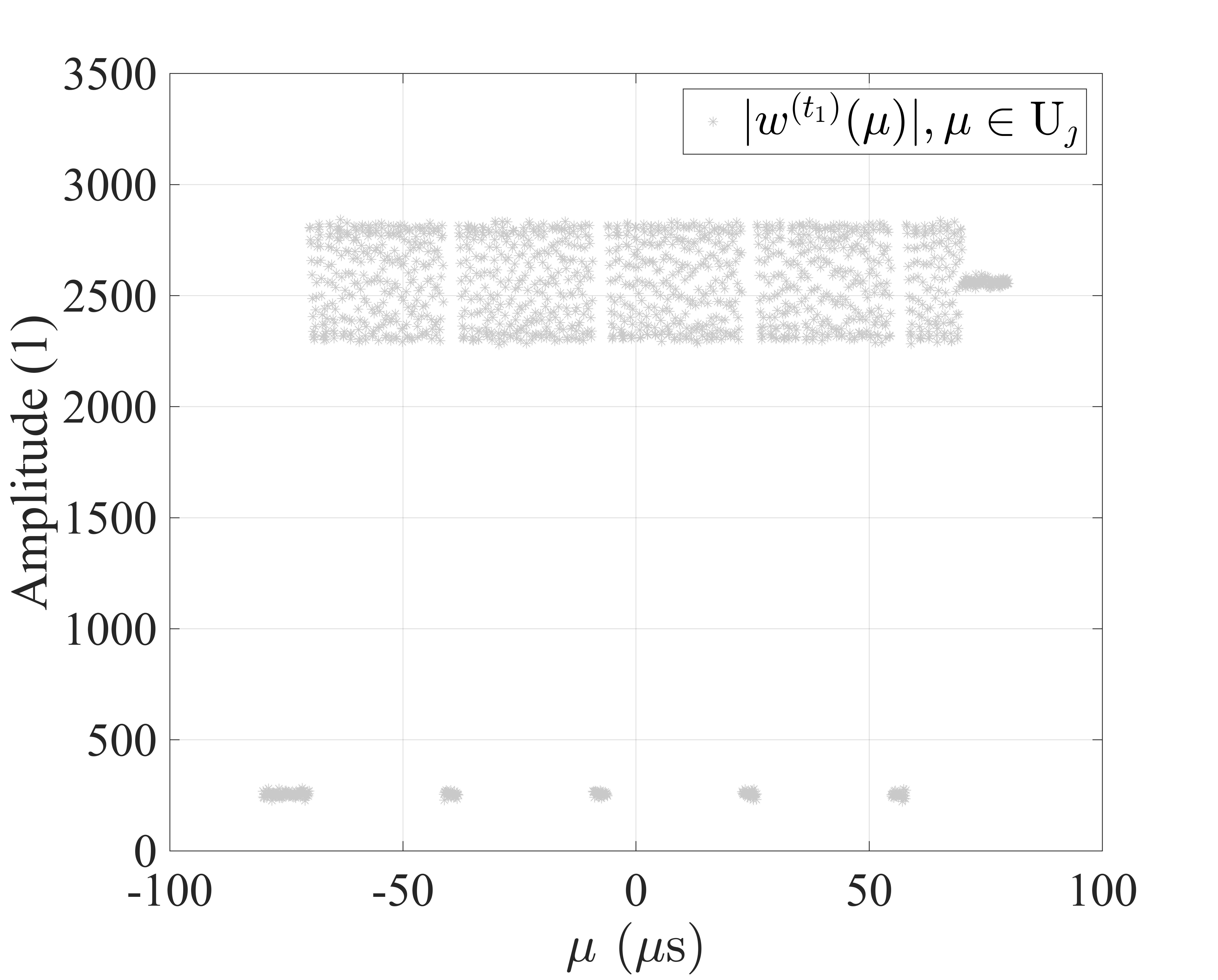}
\label{fig3(b)}}%
\subfloat[]{
\includegraphics[width=5.5 cm]{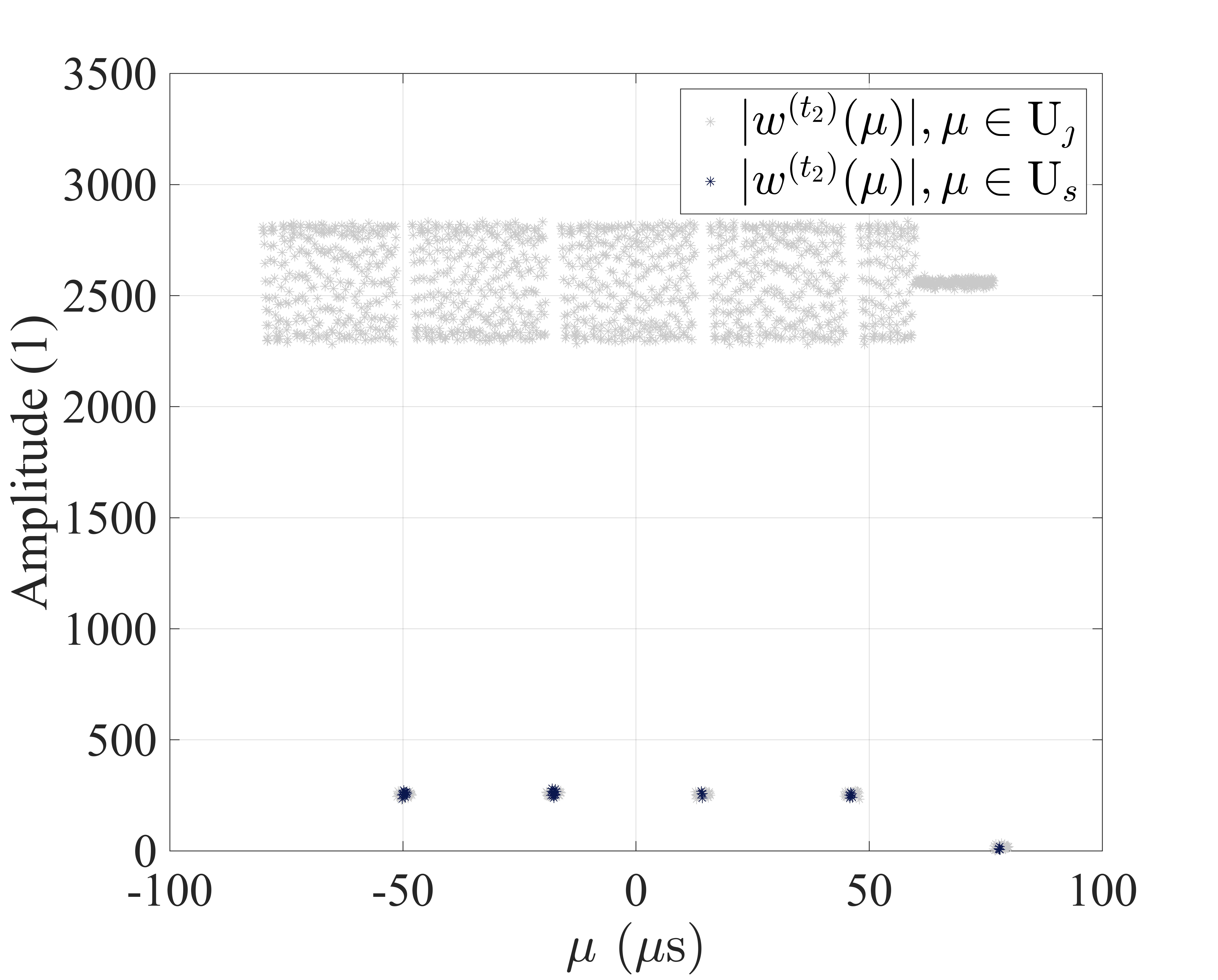}
\label{fig3(c)}}%
\\
\subfloat[]{
\includegraphics[width=5.5 cm]{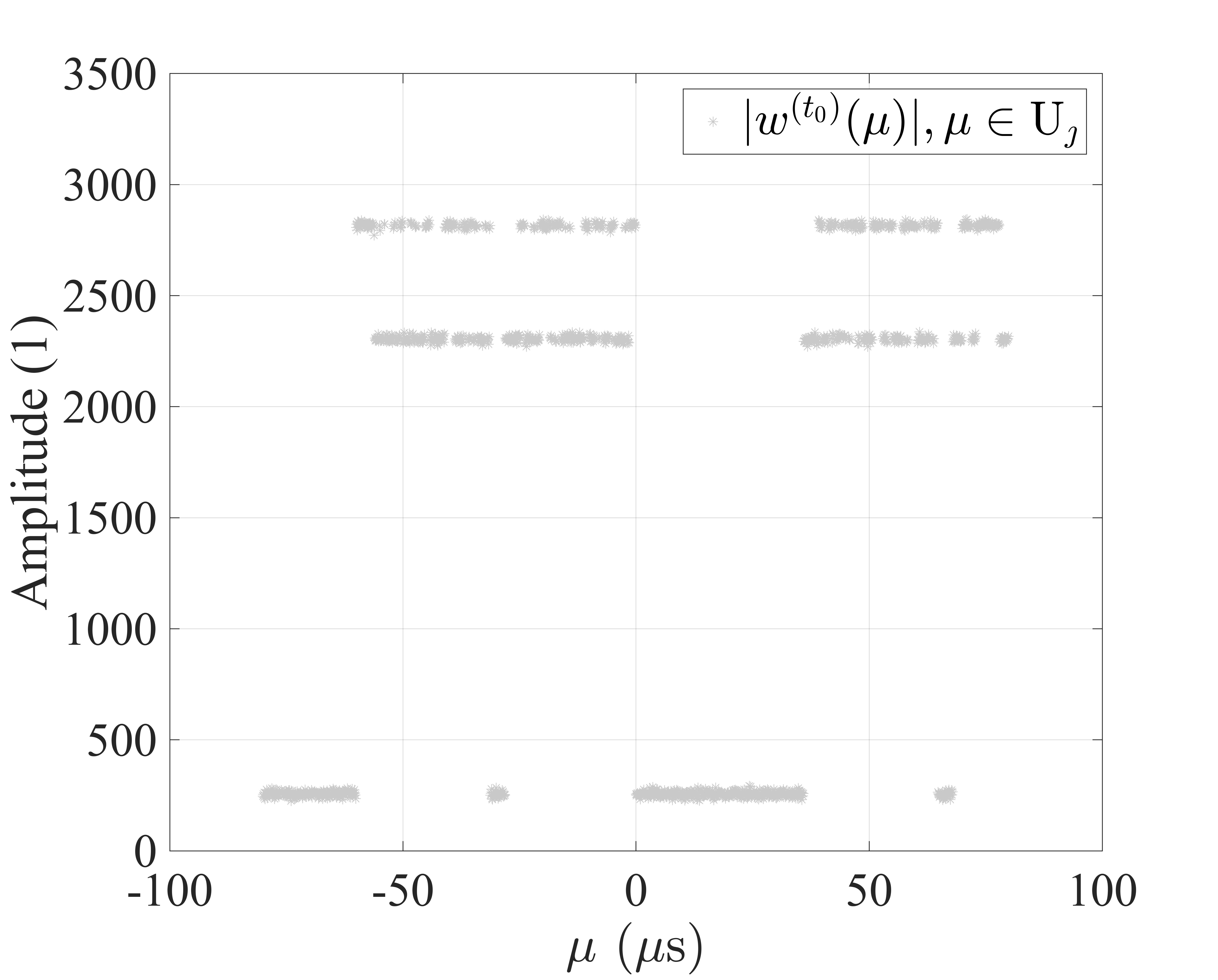}
\label{fig3(d)}}%
\subfloat[]{
\includegraphics[width=5.5 cm]{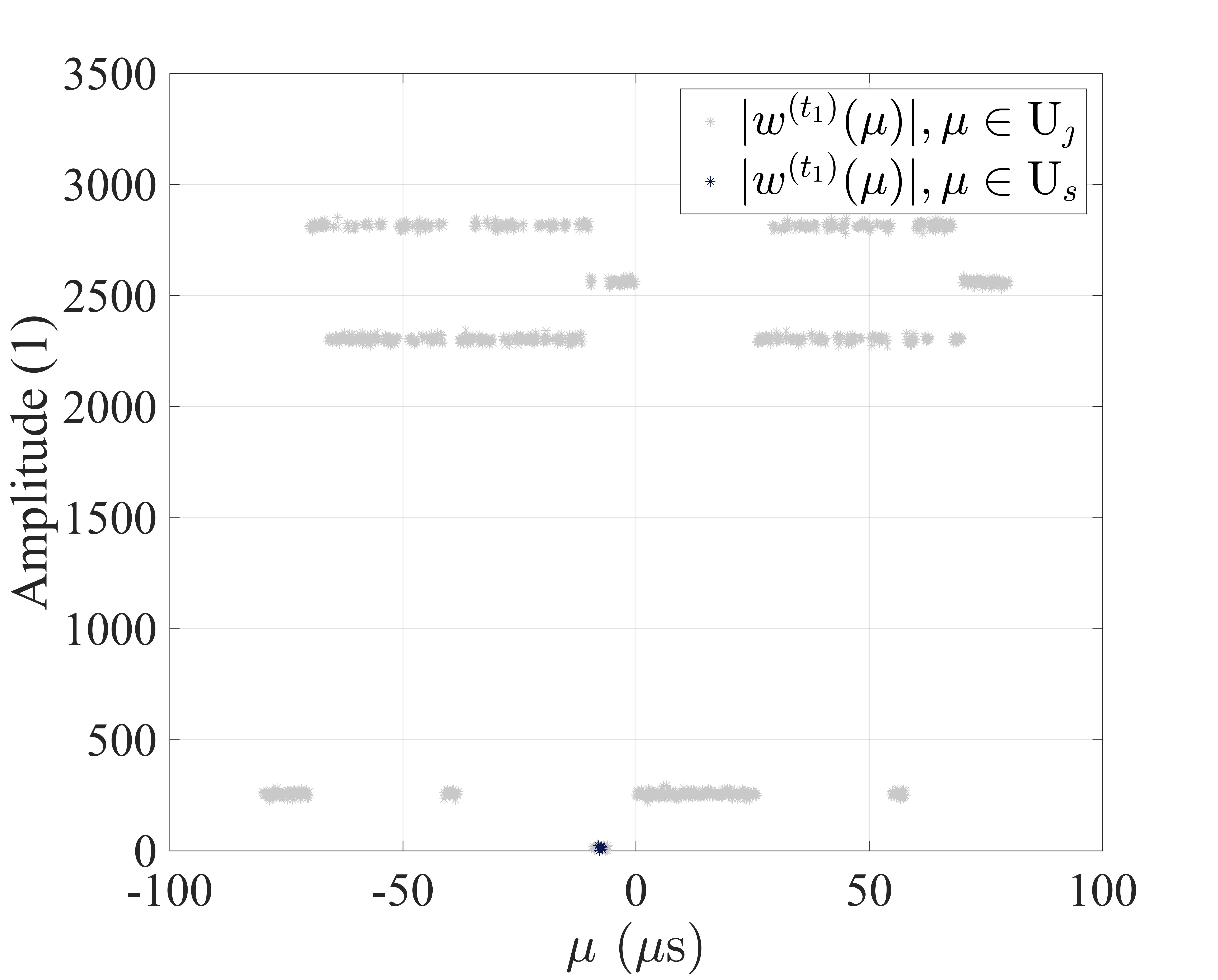}
\label{fig3(e)}}%
\subfloat[]{
\includegraphics[width=5.5 cm]{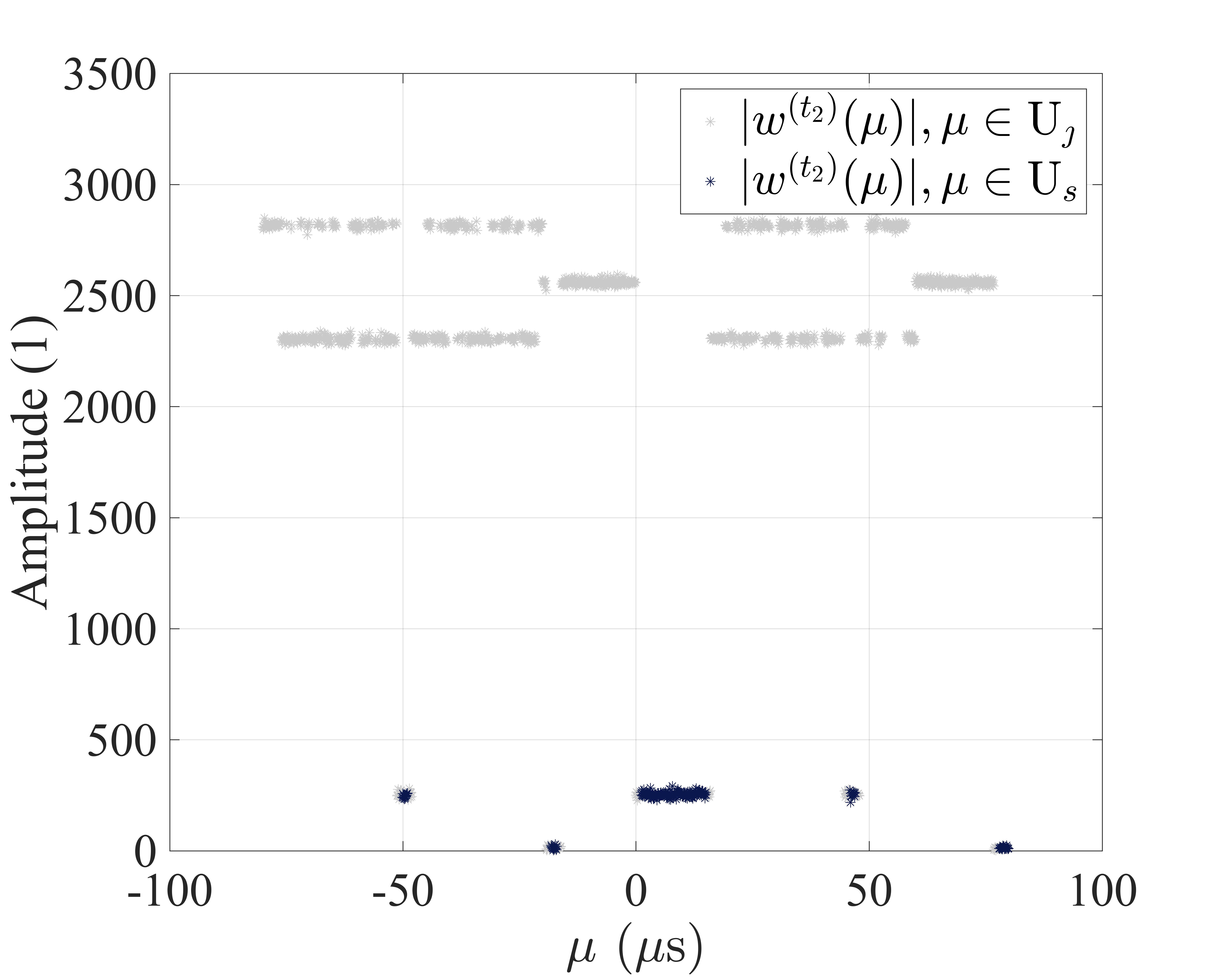}
\label{fig3(f)}}%
\\
\subfloat[]{
\includegraphics[width=5.5 cm]{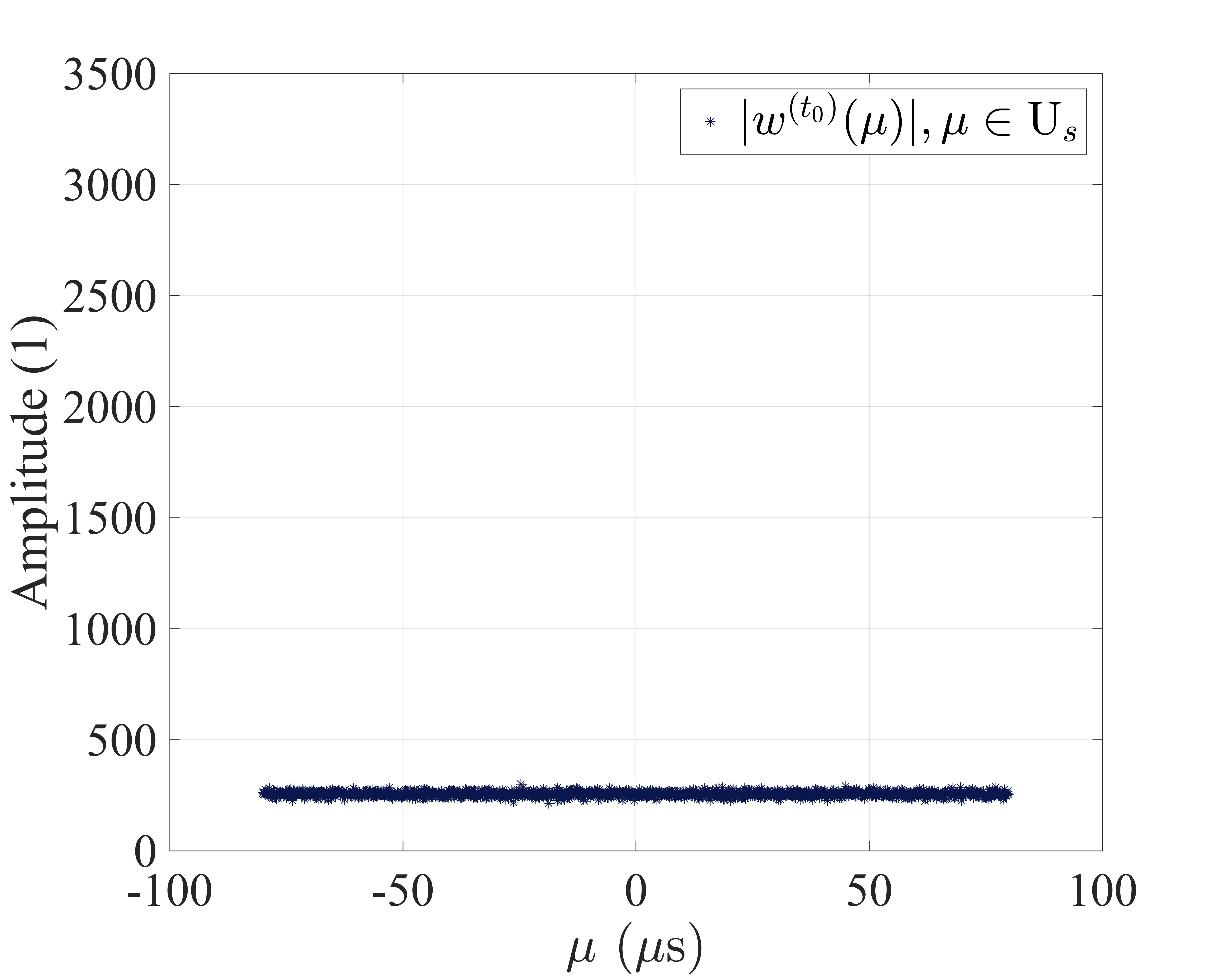}
\label{fig3(g)}}%
\subfloat[]{
\includegraphics[width=5.5 cm]{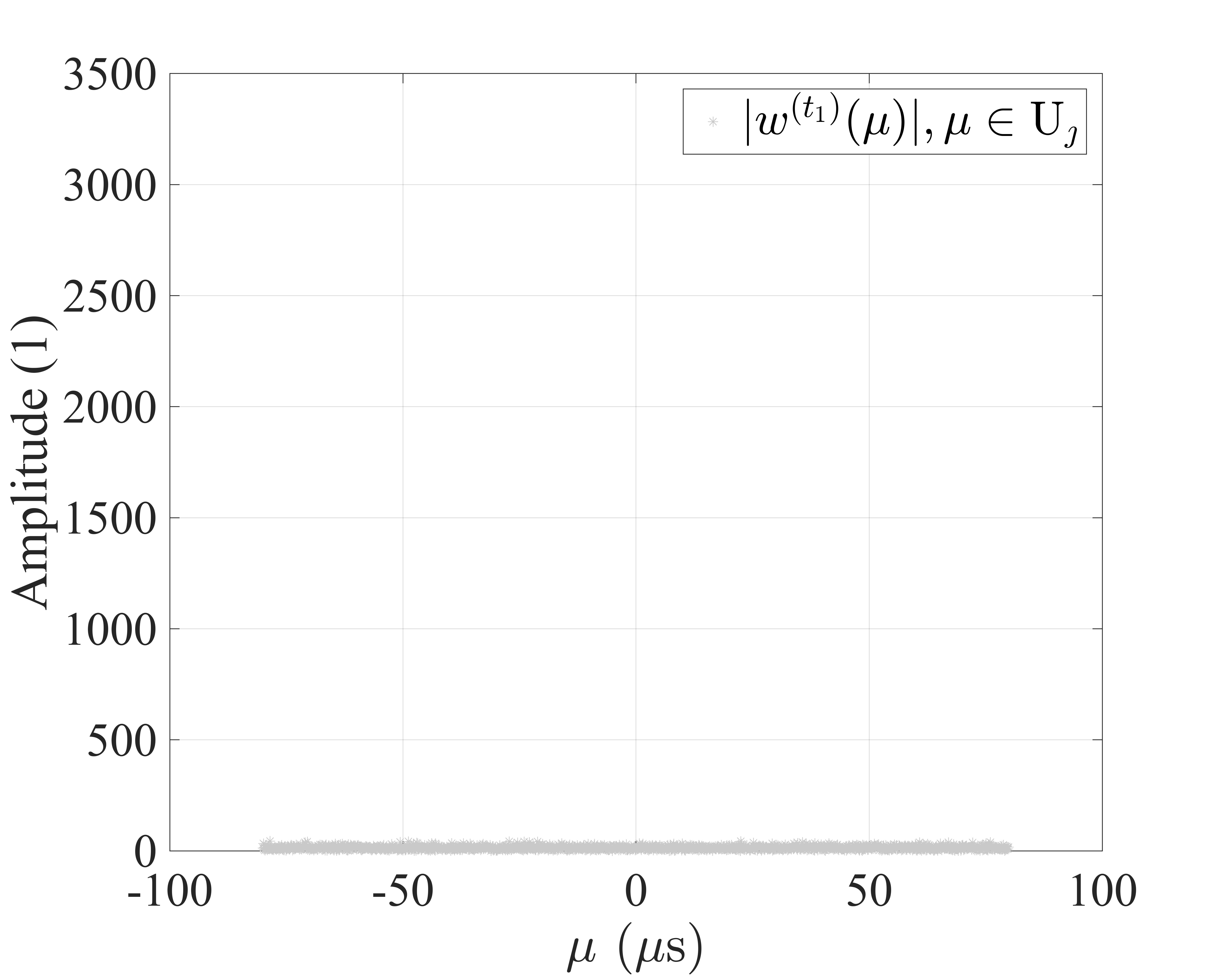}
\label{fig3(h)}}%
\subfloat[]{
\includegraphics[width=5.5 cm]{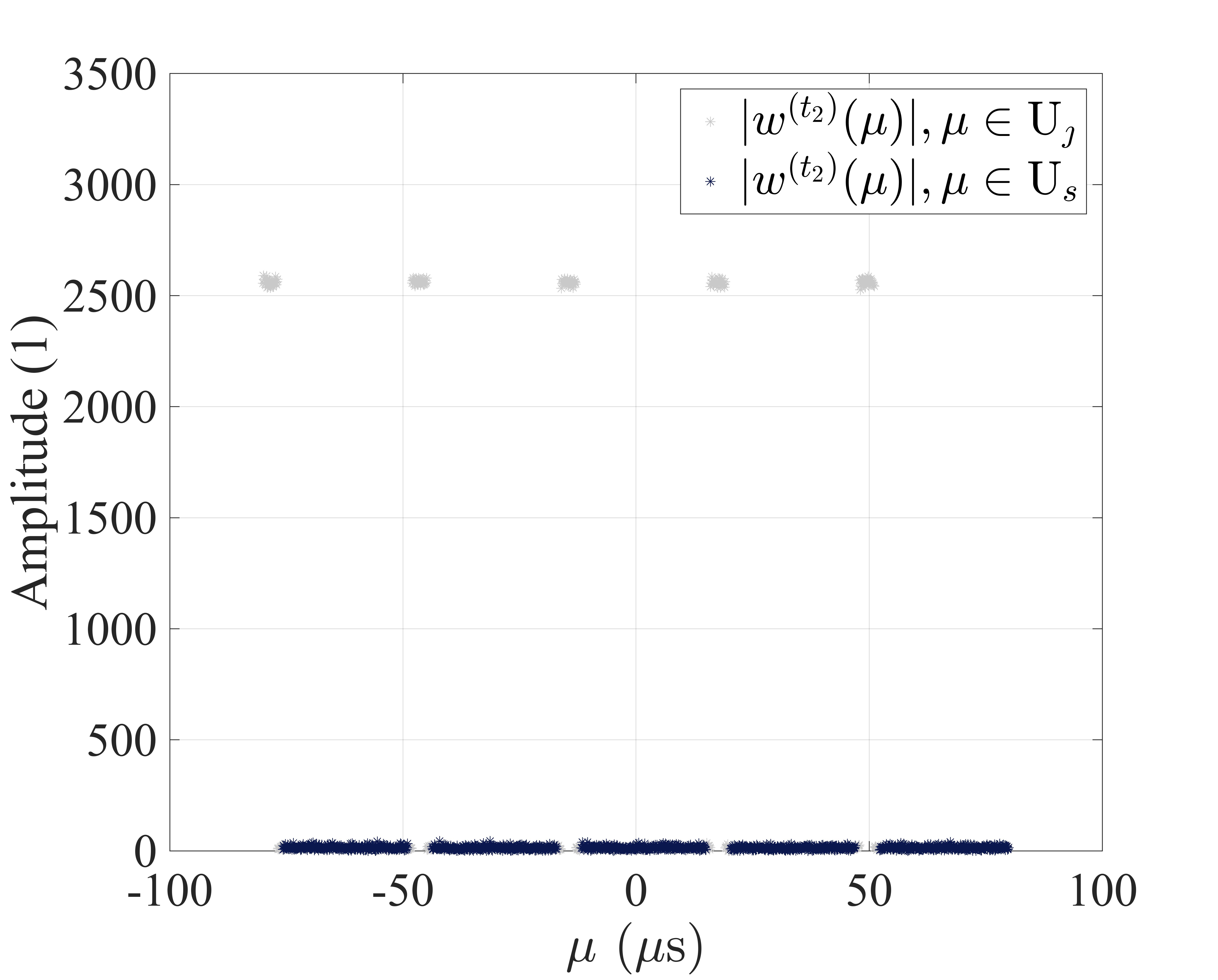}
\label{fig3(i)}}%
\centering
\caption{Coupled Results of $\mathrm{U}_s^{(t)}$ and $\mathrm{U}_\jmath^{(t)}$ with ISRRJ ($P=9$) for different waveforms. (a) LFM waveform at $t_0$; (b) LFM waveform at $t_1$; (c) LFM waveform at $t_2$. (d) Golay waveform at $t_0$; (e) Golay waveform at $t_1$; (f) Golay waveform at $t_2$. (g) WDCSS waveform at $t_0$; (h) WDCSS waveform at $t_1$; (i) WDCSS waveform at $t_2$.
\label{fig3}}
\end{figure*}
\begin{table}[htbp]
\fontsize{8}{10}\selectfont 
\caption{Simulation parameters} 
\centering
\setlength{\tabcolsep}{3pt}
\begin{tabular}{p{130pt}<{\centering}p{70pt}<{\centering}}
\toprule
Parameters&Value\\
\bottomrule
Radar carrier frequency      &$f_0 = 2$ GHz\\
Pulse repetition interval        &PRI = \SI{480}{\micro\second}\\
\hline   
Target delay        &$\tau_s =  \SI{0}{\micro\second}$\\
\hline   
Jamming delay        &$\tau_\jmath = \SI{20}{\micro\second}$\\
Repetitive repeater number &$P=9$\\
Cyclic repeater number &$Q=5$\\
\hline
Input signal to noise ratio        &SNR = $0$ dB\\
Input jamming to noise ratio        &JNR = $20$ dB\\
\hline
Coherently processing interval        &CPI =  $256$\\
\bottomrule
\end{tabular}
\label{tab1}
\end{table}
\begin{figure*}[htbp]
\centering
\subfloat[]{
\includegraphics[width=5.5 cm]{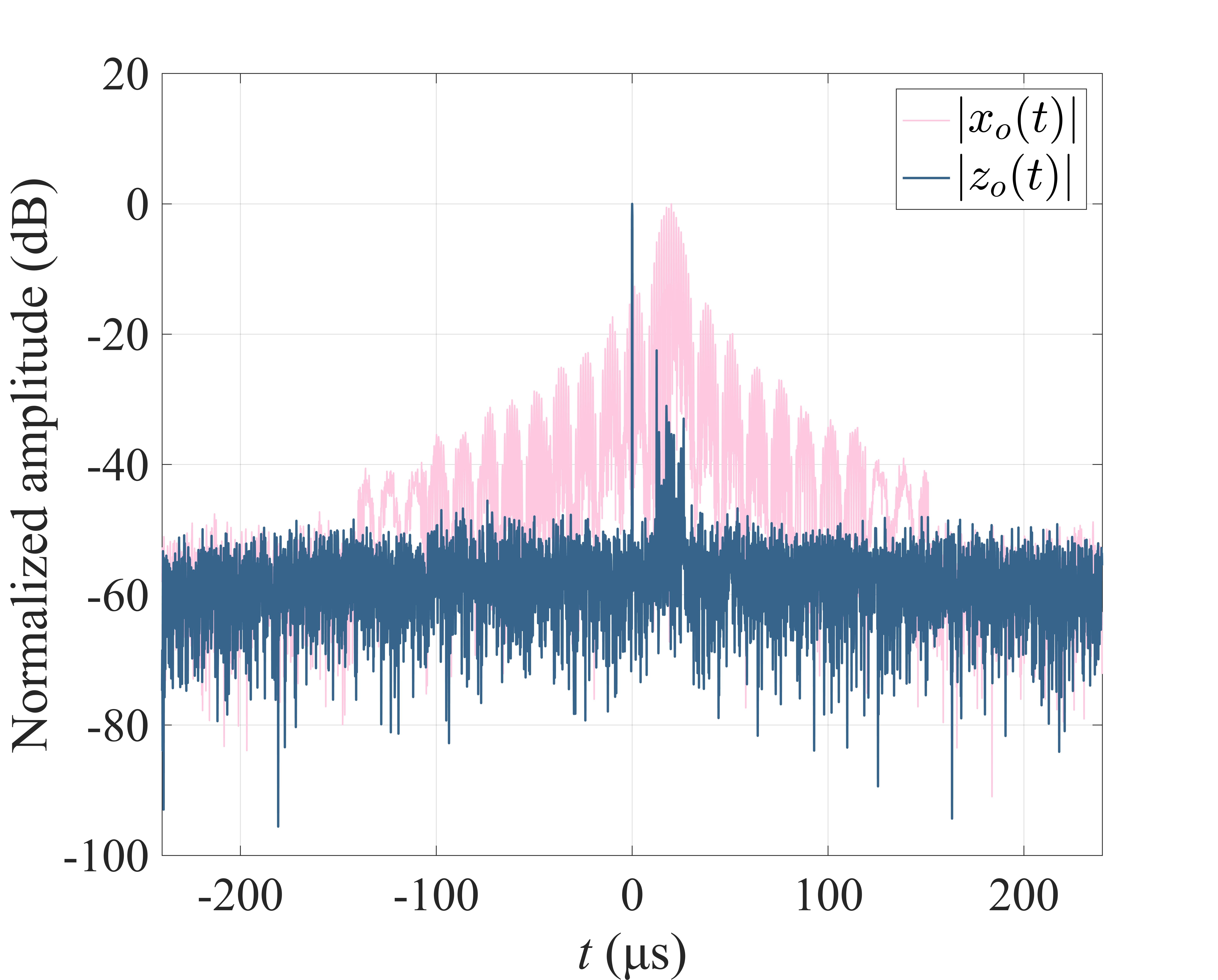}
\label{fig4(a)}}%
\subfloat[]{
\includegraphics[width=5.5 cm]{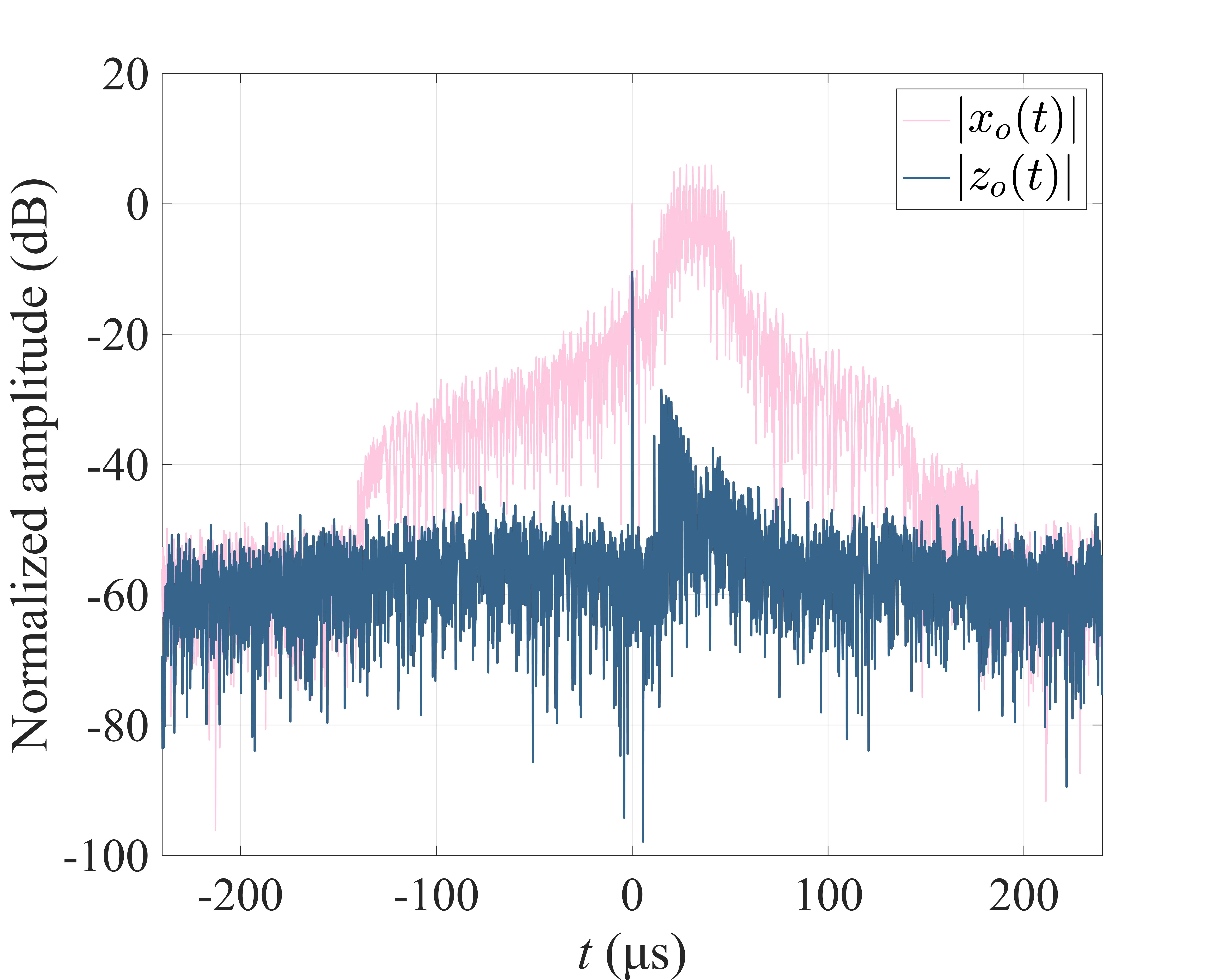}
\label{fig4(b)}}%
\subfloat[]{
\includegraphics[width=5.5 cm]{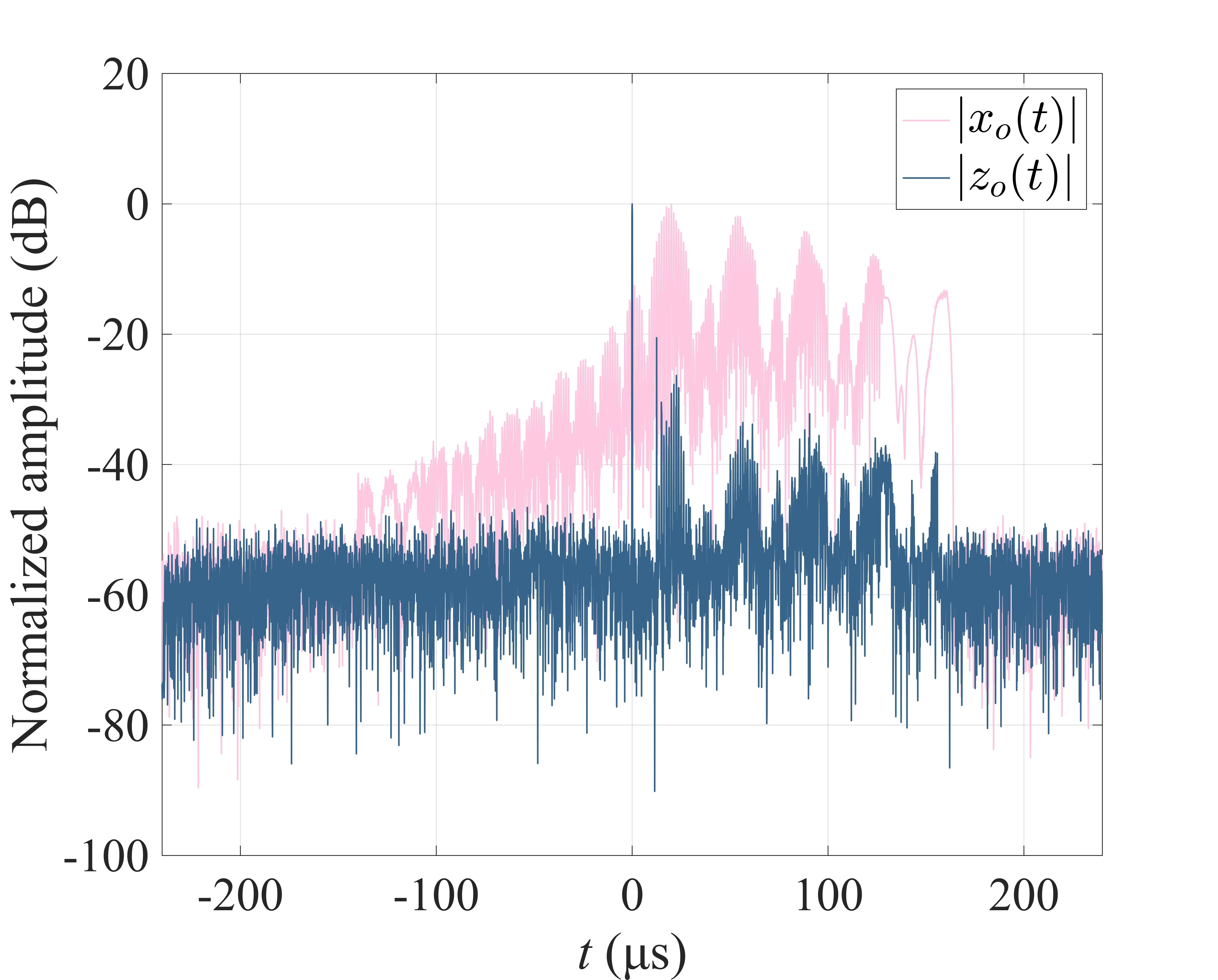}
\label{fig4(c)}}%
\\
\subfloat[]{
\includegraphics[width=5.5 cm]{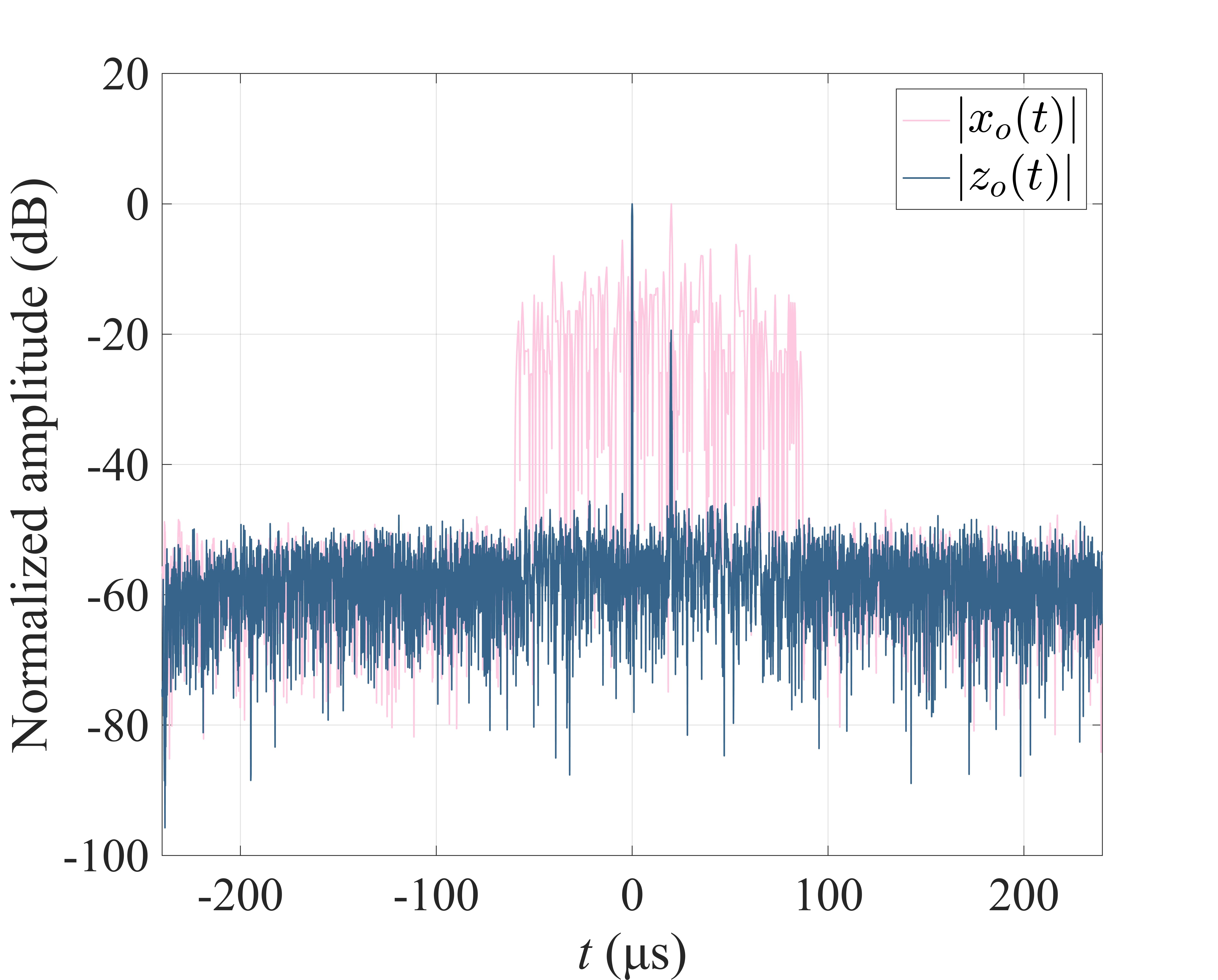}
\label{fig4(d)}}%
\subfloat[]{
\includegraphics[width=5.5 cm]{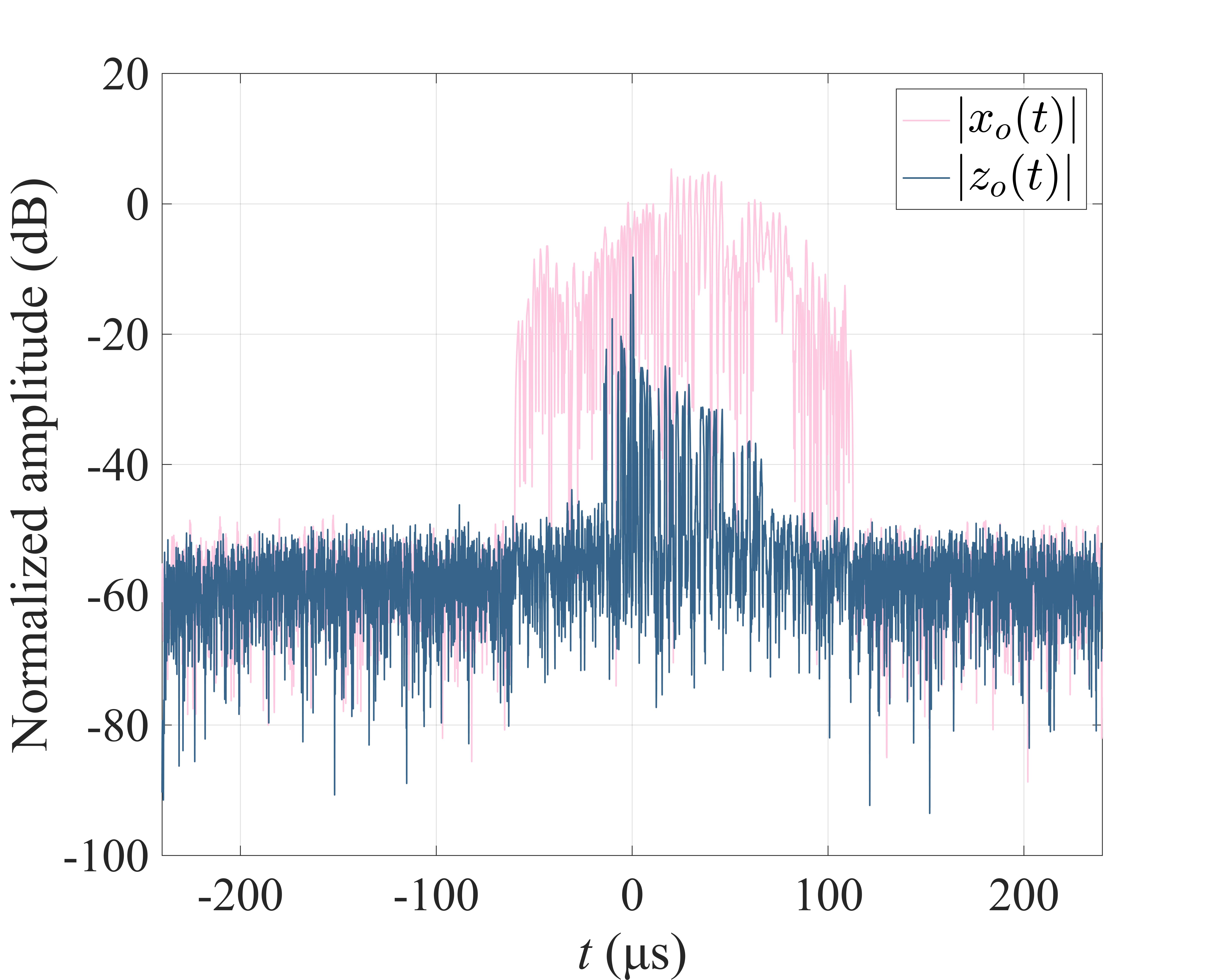}
\label{fig4(e)}}%
\subfloat[]{
\includegraphics[width=5.5 cm]{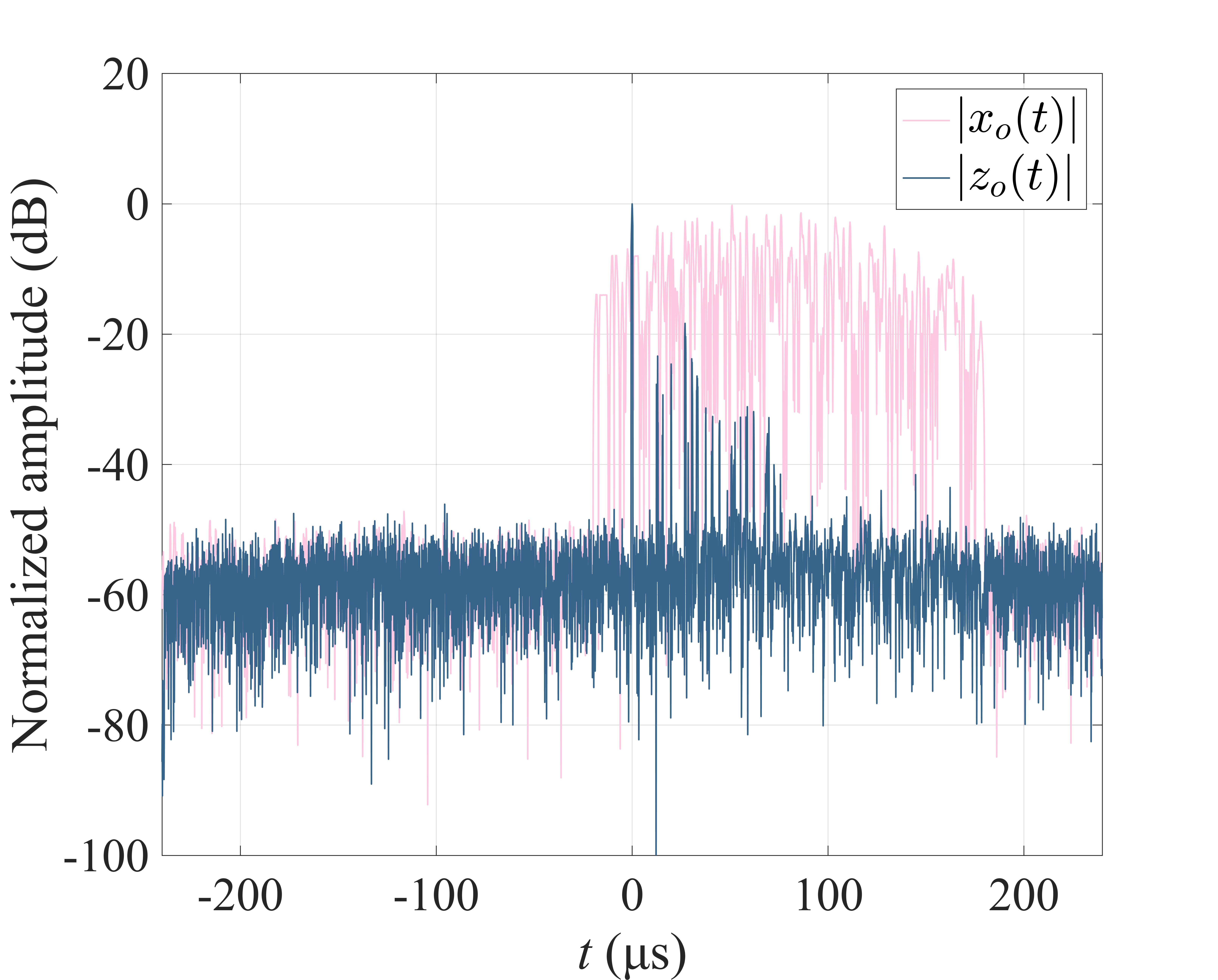}
\label{fig4(f)}}%
\\
\subfloat[]{
\includegraphics[width=5.5 cm]{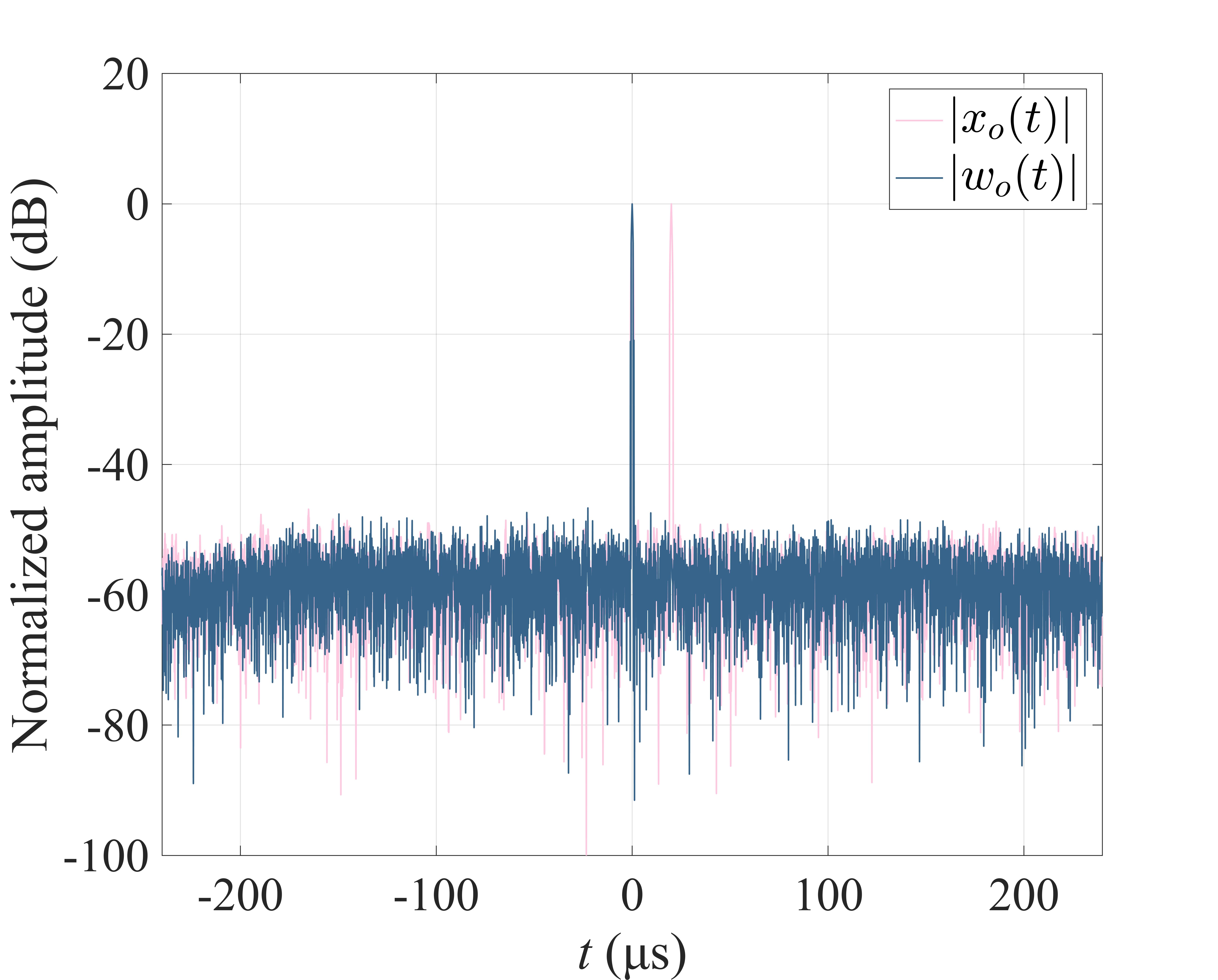}
\label{fig4(g)}}%
\subfloat[]{
\includegraphics[width=5.5 cm]{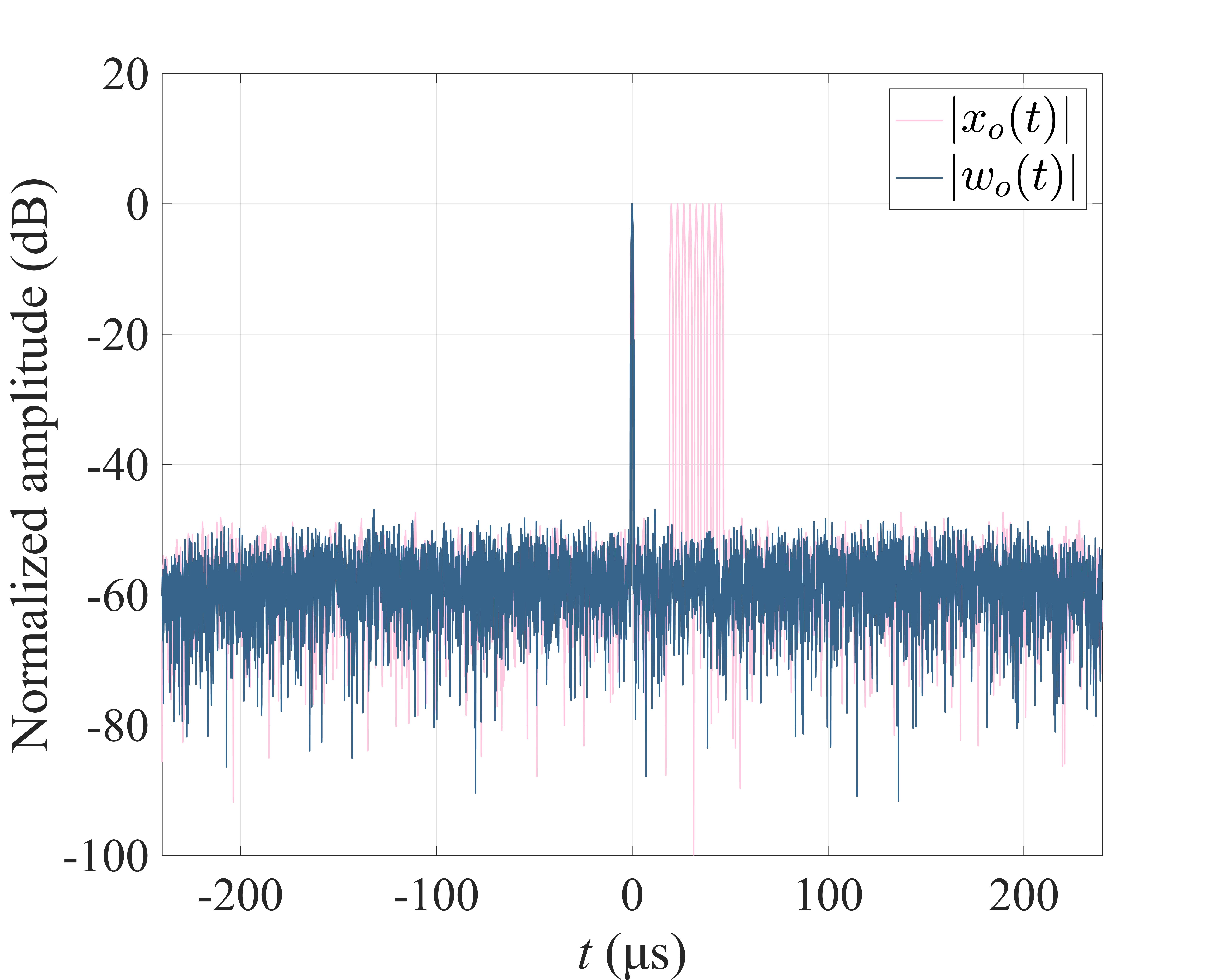}
\label{fig3(h)}}%
\subfloat[]{
\includegraphics[width=5.5 cm]{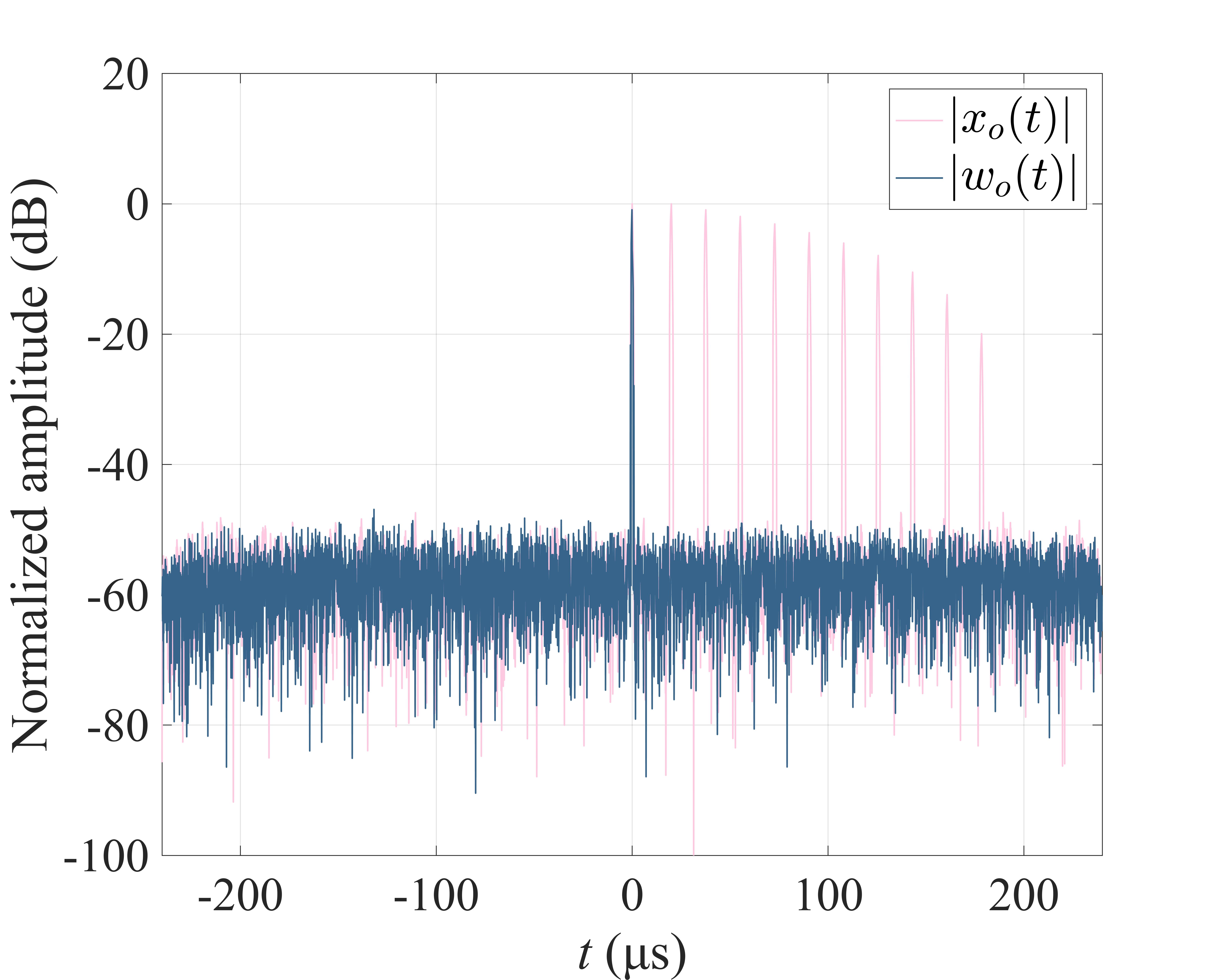}
\label{fig4(i)}}%
\centering
\caption{WD-AMF outputs for three waveform types under various ISRJ interference modes. (a) LFM waveform with ISDRJ; (b) LFM waveform with ISRRJ; (c) LFM waveform with ISCRJ. (d) Golay waveform with ISDRJ; (e) Golay waveform with ISRRJ; (f) Golay waveform with ISCRJ. (g) WDCSS waveform with ISDRJ; (h) WDCSS waveform with ISRRJ; (i) WDCSS waveform with ISCRJ.
\label{fig4}}
\end{figure*}

In order to better observe the waveform processing in the waveform domain and obtain effective and reliable results, it may be advantageous to examine the labeled outcomes of $\mathrm{U}_s^{(t)}$ and $\mathrm{U}_\jmath^{(t)}$ at several time instances. Let's define moments as $t_0 = \SI{0}{\micro\second}$, $t_1 = \SI{10}{\micro\second}$, and $t_2 = \SI{20}{\micro\second}$. At these three time instances, the labeled results for $\mathrm{U}_s^{(t)}$ and $\mathrm{U}_\jmath^{(t)}$ for the three waveform types, accompanied by an ISRRJ with $P=9$, are presented in Fig. \ref{fig3}.

Fig. \ref{fig3}(a)-Fig. \ref{fig3}(c) depict the labeled outcomes of $\mathrm{U}_s^{(t)}$ and $\mathrm{U}_{\jmath}^{(t)}$ for $w^{(t)}(\mu)$ at times $t=t_0, t_1, t_2$ when the radar transmits LFM waveforms. It is evident that $w_s^{(t)}$ and $w_\jmath^{(t)}$ manifest discernible overlapping, with the length of interfering elements $\kappa(\tau_\jmath-t)T$ in ISRRJ being non-zero, as depicted in the orange curve in Fig. \ref{fig1}(a). At these time instances, the elements of $\mathrm{U}_s^{(t)}$ are partial waveform domain elements after the removal of interference signals, and these elements are discontinuous in the waveform domain.

Fig. \ref{fig3}(d)-Fig. \ref{fig3}(f) illustrate the labeled outcomes of $\mathrm{U}_s^{(t)}$ and $\mathrm{U}_{\jmath}^{(t)}$ for $w^{(t)}(\mu)$ at times $t=t_0, t_1, t_2$ when the radar transmits Golay waveforms. Similarly, $w_s^{(t)}$ and $w_\jmath^{(t)}$ exhibit noticeable overlapping, with an interference interval length of approximately $\kappa(\tau_\jmath-t)T$, in ISRRJ being non-zero, as depicted in the orange curve in Fig. \ref{fig1}(b). At these time instances, the elements of $\mathrm{U}_s^{(t)}$ are partial waveform domain elements after the removal of interference signals, and these elements are discontinuous in the waveform domain. It is worth noting that the two sub-waveforms of the Golay complementary waveforms are orthogonal, but simulation results indicate that $w_s^{(t)}$ in the waveform domain does not possess waveform-domain complementarity because it does not meet the constraints of {\it Theorem 1}.

Fig. \ref{fig3}(g)-Fig. \ref{fig3}(i) represent the annotated results of $\mathrm{U}_s^{(t)}$ and $\mathrm{U}_{\jmath}^{(t)}$ for $w^{(t)}(\mu)$ at times $t=t_0, t_1, t_2$ during WDCSS waveform transmission by the radar. It is observed that $w_s^{(t)}$ and $w_\jmath^{(t)}$ do not demonstrate overlapping, signifying their complementarity within the waveform domain, as depicted in Fig. \ref{fig1}(c). At $t_0$, $\mathrm{U}_s^{(t)}$ spans the entire waveform domain interval, characterized by continuous elements in the waveform domain, while $\mathrm{U}_{\jmath}^{(t)}$ is the null set. At $t_1$, $\mathrm{U}_{\jmath}^{(t)}$ covers the entire waveform domain interval, whereas $\mathrm{U}_{s}^{(t)}$ remains unpopulated. At $t_2$, $\mathrm{U}_{\jmath}^{(t)}$ encompasses all the slice interference signal elements, while $\mathrm{U}_{s}^{(t)}$ solely comprises the noise elements.

Furthermore, the WD-AMF outputs for the three waveform types, accompanied by different interference modes of ISRJ, are depicted in Fig. \ref{fig4}. The gray curve represents the matched filter output $x_o(t)$, while the black curve represents the WD-AMF output $z_o(t)$ and $w_o(t)$. To facilitate the description of the anti-jamming performance of different waveforms under WD-AMF, Tab. \ref{tab2}-Tab. \ref{tab4} provides the main lobe level (MLL), sidelobe level (SLL), and peak side lobe ratio (PSLR) for various waveforms acquired using WD-AMF.

Fig. \ref{fig4}(a)-Fig. \ref{fig4}(c) illustrate the range profiles of LFM waveforms under ISDRJ, ISRRJ, and ISCRJ, respectively. The results indicate that when the interference is ISDRJ or ISCRJ, the MLL of $z_o(t)$ is equal to that of $x_o(t)$, suggesting that WD-AMF can ensure the preservation of energy in the actual target echo signal through compensation when $\eta<50\%$. However, when the interference mode is ISRRJ, the MLL of $z_o(t)$ is smaller than that of $x_o(t)$ by $10.51$ dB, indicating that even with compensation, WD-AMF cannot guarantee the preservation of energy in the actual target echo signal when $\eta>50\%$, resulting in a loss of approximately $\kappa(\tau_\jmath-t)DT$. Furthermore, upon careful observation of Fig. \ref{fig4}(a)-Fig. \ref{fig4}(c), it is noticeable that the SLL is higher than the noise floor. This is due to the discontinuity of elements in $\mathrm{U}_s^{(t)}$, as shown in Fig. \ref{fig3}(a)-Fig. \ref{fig3}(c), leading to an increase in the SLL during signal mismatch.

Fig. \ref{fig4}(d)-Fig.\ref{fig4}(f) depict the range profiles of Golay waveforms under ISDRJ, ISRRJ, and ISCRJ, respectively. The WD-AMF results exhibit a similar performance to LFM waveforms, with the distinction being that the side lobe level of WD-AMF is higher than that of LFM waveforms. This is due to the time-domain complementary properties of Golay being sensitive to Doppler effects, and the discontinuous integration elements of $w^{(t)}(\mu), \mu\in\mathrm{U}_s^{(t)}$ disrupt this complementary nature, resulting in an increase in SLL.

Fig. \ref{fig4}(g)-Fig. \ref{fig4}(i) illustrate the range profiles of WDCSS waveforms under ISDRJ, ISRRJ, and ISCRJ, respectively. The results show that, regardless of the interference mode, the MLL of $w_o(t)$ is consistently equal to that of $x_o(t)$, and the SLL of $w_o(t)$ is on par with the noise level. This is because WDCSS waveforms exhibit waveform-domain complementarity, enabling $w_s^{(t)}$ and $w_\jmath^{(t)}$ to be treated as two single-component signals at different time instances, as shown in Fig. \ref{fig3}(g)-Fig. \ref{fig3}(i). Therefore, WDCSS waveforms are the most suitable anti-jamming waveforms for WD-AMF.
\begin{figure}[htbp]
\centering
\subfloat[]{
\includegraphics[width=6 cm]{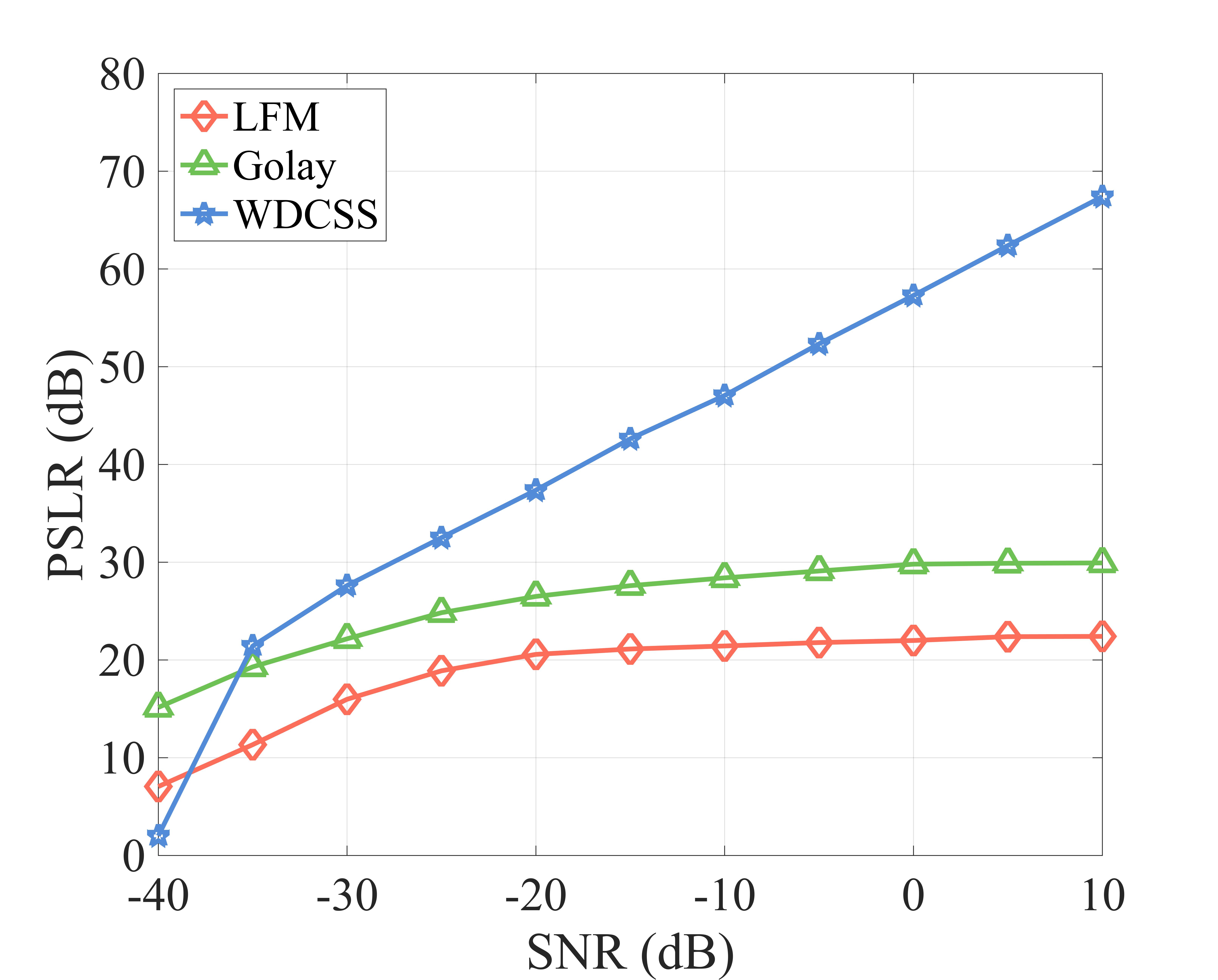}
\label{fig5(a)}}%
\\
\subfloat[]{
\includegraphics[width=6 cm]{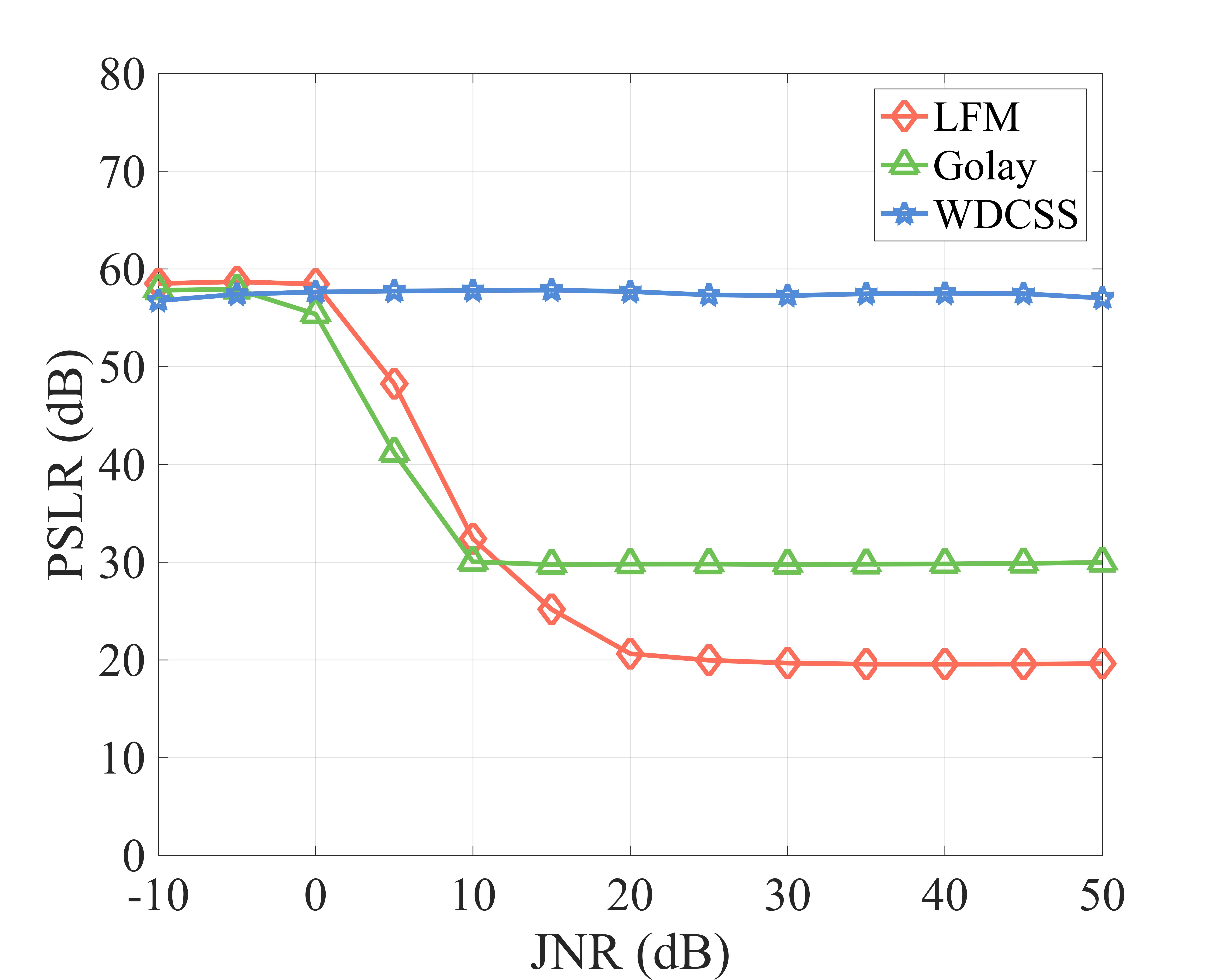}
\label{fig5(b)}}%
\centering
\caption{Variations in PSLR of WD-AMF outputs under different SNR and JNR conditions, with a specific focus on ISRRJ conditions. (a) Curves depicting PSLR variation with SNR; (b) Curves illustrating PSLR variation with JNR.
\label{fig5}}
\end{figure}

From the above analysis, it can be inferred that different waveforms exhibit varying waveform domain complementarity, leading to distinct PSLR values in their WD-AMF performance, particularly under ISRRJ conditions. Therefore, it is imperative to further validate the WD-AMF performance of different waveforms in the context of Tab. \ref{tab1} scenarios under various input SNR and input JNR conditions, particularly in the presence of ISRRJ. To mitigate the stochastic effects introduced by noise, $500$ Monte Carlo simulations were conducted for each input SNR and input JNR value. Fig. \ref{fig5} represents the average PSLR results obtained from multiple simulations. It is noteworthy that for Fig. \ref{fig5}(a), the input JNR is fixed at JNR = $20$ dB, while for Fig. \ref{fig5}(b), the input SNR is fixed at SNR = $0$ dB.
\begin{table}[htbp]
\fontsize{8}{10}\selectfont 
\caption{WD-AMF outputs of the LFM waveform} 
\centering
\setlength{\tabcolsep}{3pt}
\begin{tabular}{p{50pt}<{\centering}p{50pt}<{\centering}p{50pt}<{\centering}p{50pt}<{\centering}}
\toprule
ISRJ mode&MLL (dB) $(t=\tau_s)$ & SLL (dB) $(t=\tau_\jmath)$ & PSLR (dB)\\
\bottomrule
ISDRJ      &$0$  &$-35.34$   &$35.34$ \\
ISRRJ      &$-10.51$  &$-30.40$   &$19.89$\\
ISCRJ        &$0$ & $-29.09$& $29.09$\\
\bottomrule
\end{tabular}
\label{tab2}
\end{table}
\begin{table}[htbp]
\fontsize{8}{10}\selectfont 
\caption{WD-AMF outputs of the Golay waveform} 
\centering
\setlength{\tabcolsep}{3pt}
\begin{tabular}{p{50pt}<{\centering}p{50pt}<{\centering}p{50pt}<{\centering}p{50pt}<{\centering}}
\toprule
ISRJ mode&MLL (dB) $(t=\tau_s)$ & SLL (dB) $(t=\tau_\jmath)$ & PSLR (dB)\\
\bottomrule
ISDRJ      &$0$  &$-26.01$   &$26.01$ \\
ISRRJ        &$-57.00$ & $-33.25$& $-23.75$\\
ISCRJ        &$0$ & $-26.42$& $26.42$\\
\bottomrule
\end{tabular}
\label{tab3}
\end{table}
\begin{table}[htbp]
\fontsize{8}{10}\selectfont 
\caption{WD-AMF outputs of the WDCSS waveform} 
\centering
\setlength{\tabcolsep}{3pt}
\begin{tabular}{p{50pt}<{\centering}p{50pt}<{\centering}p{50pt}<{\centering}p{50pt}<{\centering}}
\toprule
ISRJ mode&MLL (dB) $(t=\tau_s)$ & SLL (dB) $(t=\tau_\jmath)$ & PSLR (dB)\\
\bottomrule
ISDRJ      &$0$  &$-55.62$   &$55.62$ \\
ISRRJ        &$0$ & $-52.46$& $52.46$\\
ISCRJ        &$0$ & $-53.58$& $53.58$\\
\bottomrule
\end{tabular}
\label{tab4}
\end{table}

From the simulation results in Fig. \ref{fig5}(a), it can be observed that when $\mathrm{SNR}>-20$ dB, the PSLR of LFM and Golay waveforms remains relatively constant as SNR increases. This is because their WD-AMF output's SLL is greater than the noise output level. However, when $\mathrm{SNR}<-20$ dB, the SLL becomes smaller than the noise output level, resulting in a nearly linear increase in PSLR with rising SNR. Similarly, it is noted that when $\mathrm{SNR}>-35$ dB, the PSLR of the WDCSS waveform linearly increases with SNR. This is attributed to the complementary nature of the WDCSS waveform in the waveform domain, causing its WD-AMF output's SLL to match the noise output level. Conversely, when $\mathrm{SNR}<-35$ dB, the PSLR experiences a steep decline as SNR decreases. This is because the WDCSS waveform also exhibits time-domain complementarity, leading to the $\hat E^{(\tau_s)}$ of the real target echo not benefiting from the sidelobe gain of ISRJ, resulting in a smaller $\hat E^{(\tau_s)}$ at low SNR. Consequently, $\mathrm{U}_s^{(\tau_s)}$ loses some elements in the waveform domain, leading to a decrease in MLL and PSLR. As anticipated, the WDCSS waveform exhibits higher PSLR, indicating superior interference resistance performance.

From Fig. \ref{fig5}(b), it can be observed that when $\mathrm{JNR}<0$ dB, the PSLR of LFM and Golay waveforms remains relatively constant as SNR increases. This is because with a small JNR, $A_\jmath+A_s<\hat E^{(\tau_s)}$, and in such cases, $\mathrm{U_s}^{(\tau_s)}$ covers the entire waveform domain, thus incurring no MLL loss. Conversely, when $\mathrm{JNR}>0$ dB, the PSLR of LFM and Golay waveforms gradually decreases as SNR increases, and then remains constant. This is because with $A_\jmath+A_s>\hat E^{(\tau_s)}$, MLL and PSLR suffer losses. For the WDCSS waveform, due to its complementary nature in the waveform domain, the real target echo signal and ISRJ do not exhibit overlapping in the waveform domain. Therefore, its PSLR is always equivalent to MLL. Consequently, the performance of the WDCSS waveform in WD-AMF is unaffected by JNR.

In conclusion, based on the above analysis, in the simulation scenarios as presented in Tab. \ref{tab1}, for the WDCSS waveform, under all parameter configurations, PSLR exceeds 18 dB when $\mathrm{SNR}>-35$ dB, thus meeting the detection requirements of this scenario.

\subsection{Parameter sensitivity analysis}

In the preceding section, we observed that the PSLR of WD-AMF output for WDCSS waveforms remains unaffected by the ISRJ parameter $\eta$ due to their waveform-domain complementarity. However, for LFM and Golay waveforms, which lack such complementarity, the PSLR of their WD-AMF outputs is significantly influenced by $\eta$. To investigate the impact of this critical ISRJ parameter $\eta$ on PSLR for different waveforms, this section employs ISRRJ as an example and conducts simulation and comparative experiments by adjusting $\eta$.

\begin{figure}[htbp]
\includegraphics[width=6 cm]{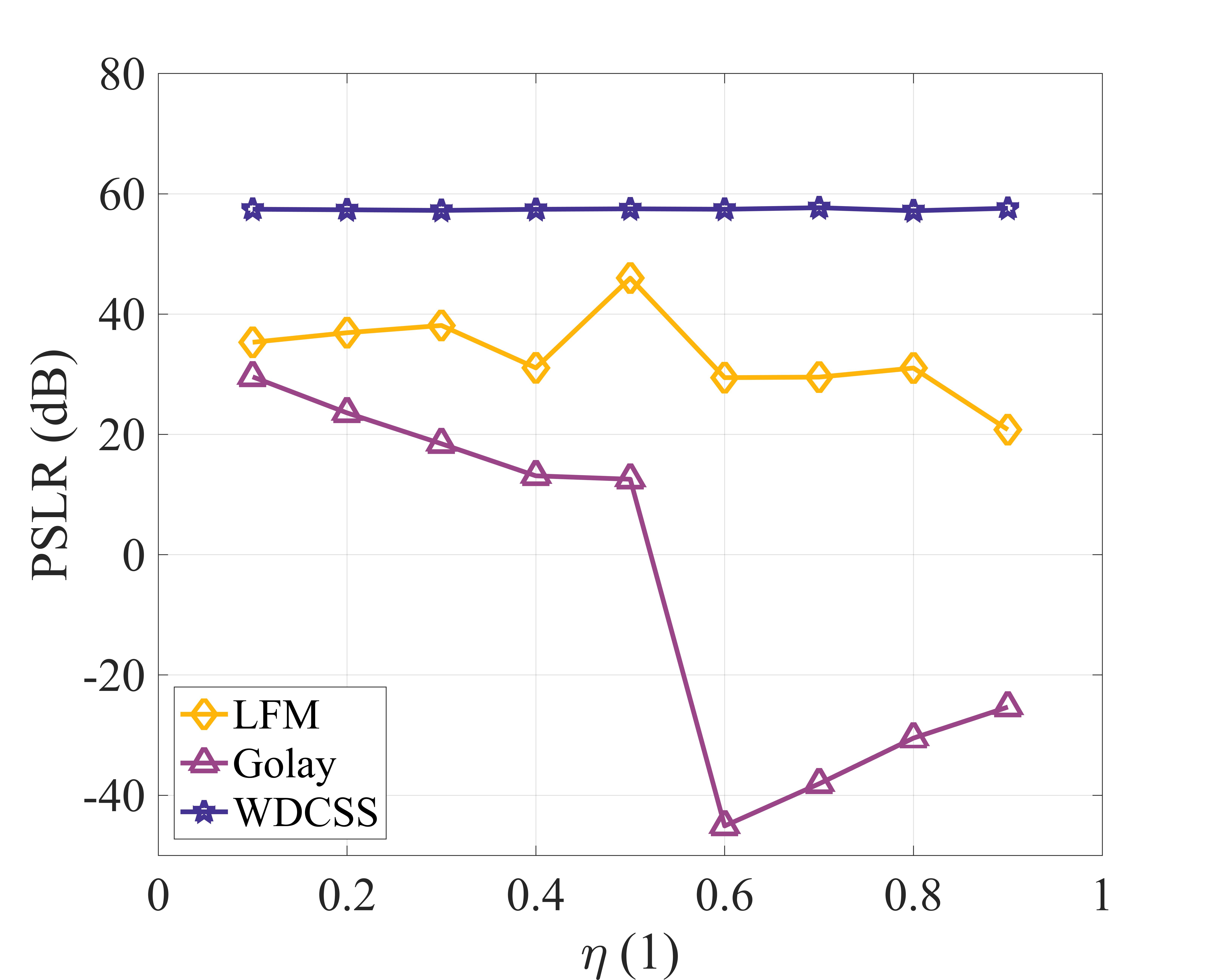}
\centering
\caption{Variations in PSLR of WD-AMF outputs under different $\eta$ conditions, with a specific focus on ISRRJ conditions. 
\label{fig6}}
\end{figure}

We will maintain the simulation parameters as outlined in Tab. \ref{tab1}, while varying the range of repetitions, denoted as $P\in[1,9]$, and the corresponding duty cycle $\eta\in[0.1,0.9]$. Fig. \ref{fig6} illustrates the variation in the PSLR of WD-AMF output for the three waveforms with respect to the changes in $\eta$.

Simulation results indicate that the PSLR of the WDCSS waveform remains consistently at its maximum output SNR across the entire range of parameter variation for $\eta$. This implies that the WDCSS waveform is insensitive to changes in $\eta$. Conversely, the PSLR of LFM and Golay waveforms exhibits significant fluctuations with variations in $\eta$ due to changes in MLL and the fluctuation of SLL. As anticipated, the WDCSS waveform, possessing waveform-domain complementarity, demonstrates superior and robust interference resistance performance. Furthermore, all PSLR values for the WDCSS waveform are above 55 dB, further validating the practicality of the proposed method in engineering applications.

\section{CONCLUSION}

This paper introduces a waveform-domain complementary signal set (the WDCSS waveform), to address the issue of ISRJ resistance. Through an analysis of ISRJ and the waveform response functions of the transmitted waveforms, it is revealed that waveform-domain non-sparsity is the primary factor causing the degradation of interference resistance performance in waveform-domain adaptive matched filtering (WD-AMF) when there is signal mismatch. Improved WD-AMF, adapted to the WDCSS waveform, is also considered because the WDCSS waveform does not necessitate additional waveform-domain signal compensation. The anti-ISRJ problem is formulated, and the WDCSS waveform is designed by introducing the Walsh-Hadamard matrix.

Several simulations are conducted to demonstrate the effectiveness of the proposed method. Simulation results indicate that WD-AMF using the WDCSS waveform achieves superior anti-ISRJ performance compared to WD-AMF using separately designed waveforms. Parametric sensitivity analysis shows that the WDCSS waveform is insensitive to the signal duty cycle, regardless of the interference mode.

Since both the design and processing of WD-AMF with the WDCSS waveform do not require prior information about ISRJ-related parameters, the transmitter only needs a single operating mode to achieve ISRJ suppression for different scenarios. For future work, attention may be directed towards the rapid implementation of WD-AMF and the design of WDCSS waveforms with a larger Doppler tolerance, enabling the proposed method to be applied to real-time target tracking for ISRJ suppression and improving anti-ISRJ performance for high-speed moving targets.

\bibliographystyle{unsrt}
\bibliography{reference}

\begin{IEEEbiographynophoto}{Hanning Su} received the B.Sc degree in electronic engineering from Xidian University in 2018. He is currently working towards the Ph.D. degree in signal and information processing with the National Key Lab of Science and Technology on ATR, National University of Defense Technology. His current research interests  include radar signal processing, target tracking, and radar anti-jamming.
\end{IEEEbiographynophoto}
\begin{IEEEbiographynophoto}{Qinglong Bao} received his B.Sc and Ph.D degrees from the National University of Defense Technology, Changsha, China, in 2003 and 2010, respectively. Currently, he is an Associate Professor with the School of Electronic Science, National University of Defense Technology. His current research interests include radar data acquisition and signal processing.
\end{IEEEbiographynophoto}
\begin{IEEEbiographynophoto}{Jiameng Pan} received the B.E. degree in Zhejiang University in 2013, and the Ph.D. degree in National University of Defense Technology in 2020. He is currently a lecturer with the College of Electronic Science and Technology, National University of Defense Technology. His main research interests include radar signal processing, target tracking, and radar anti-jamming.
\end{IEEEbiographynophoto}
\begin{IEEEbiographynophoto}{Fucheng Guo} received the Ph.D. degree in information and communication engineering from the National University of Defense Technology (NUDT), Changsha, Hunan, China, in 2002.,He is now a Professor in the School of Electronic Science, NUDT. His research interests include source localization, target tracking, and radar/communication signal processing.
\end{IEEEbiographynophoto}
\begin{IEEEbiographynophoto}{Weidong Hu} was born in September 1967. He received the B.S. degree in microwave technology and the M.S. and Ph.D. degrees in communication and electronic system from the National University of Defense Technology, Changsha, China, in 1990, 1994, and 1997, respectively.

He is currently a Full Professor in the ATR Laboratory, National University of Defense Technology, Changsha. His research interests include radar signal and data processing.
\end{IEEEbiographynophoto}

\end{document}